\documentclass[12pt,letterpaper]{article}
\usepackage[includeheadfoot,
            marginratio={1:1,2:3},
            width=412pt,
            height=688pt,]{geometry}

\usepackage{amsmath}
\usepackage{amsfonts}
\usepackage{amssymb}
\usepackage{stmaryrd}
\usepackage{empheq}

\newcommand{\nc}{\newcommand}
\nc{\gl}{\llbracket}
\nc{\gr}{\rrbracket}

\newcommand{\eq}[1]{\begin{equation}
                     \begin{split} #1 \end{split}
                     \end{equation}}

\newcommand{\fa}{\hat}

\newcommand{\fc}{\tilde }
\newcommand{\Lie}{{\cal L}}

\allowdisplaybreaks[2]
\numberwithin{equation}{section}

\begin{document}

\vspace*{-1.5cm}
\begin{flushright}
  {\small
  MPP-2012-147\\
  DFPD-2012-TH-14

  }
\end{flushright}

\vspace{1.5cm}
\begin{center}
{\LARGE
Non-geometric strings, symplectic gravity  and \\[0.3cm]
differential geometry of Lie algebroids
}
\vspace{0.2cm}

\end{center}

\vspace{0.35cm}
\begin{center}
  Ralph Blumenhagen$^{1}$, Andreas Deser$^{1}$, Erik Plauschinn$^{2,3}$ and
Felix Rennecke$^{1}$
\end{center}

\vspace{0.1cm}
\begin{center}
\emph{$^{1}$ Max-Planck-Institut f\"ur Physik (Werner-Heisenberg-Institut) \\
   F\"ohringer Ring 6,  80805 M\"unchen, Germany } \\[0.1cm]
\vspace{0.25cm}
\emph{$^{2}$ Dipartimento di Fisica e Astronomia ``Galileo Galilei'' \\
Universit\`a  di Padova, Via Marzolo 8, 35131 Padova, Italy}  \\[0.1cm]
\vspace{0.25cm}
\emph{$^{3}$ INFN, Sezione di Padova \\
Via Marzolo 8, 35131 Padova, Italy}  \\

\end{center}

\vspace{1cm}


\begin{abstract}
Based on the structure of a Lie algebroid for non-geometric fluxes
in string theory, a differential-geometry calculus is developed 
which combines usual
diffeomorphisms with so-called $\beta$-diffeomorphisms emanating from
gauge symmetries of the Kalb-Ramond field.
This allows to construct a bi-invariant action of Einstein-Hilbert type
comprising a metric, a \mbox{(quasi-)}sym\-plec\-tic structure $\beta$ and a dilaton.
As a salient feature, this symplectic gravity action 
and the resulting equations of motion
take a form which is similar to the standard action and field equations.
Furthermore, the two actions turn out to be related via a field redefinition 
reminiscent of the Seiberg-Witten limit.
Remarkably, this redefinition admits a direct
generalization to higher-order
$\alpha'$-corrections and to the additional fields and couplings appearing
in the effective action of the superstring.
Simple solutions to the equations of motion of the symplectic gravity action, including
Calabi-Yau geometries, are discussed.
\end{abstract}

\clearpage

\tableofcontents


\section{Introduction}
\label{sec:intro}

In the conventional approach to string theory
the universal massless excitations are
expressed in terms of a metric, a Kalb-Ramond field  and a dilaton.
Their dynamics, at leading order in a large
distance expansion, can be determined from string scattering amplitudes
and is found to be described
by an extension of the Einstein-Hilbert action
\eq{
\label{stringaction_intro}
S= \frac{1}{2\hspace{0.5pt}\kappa^2}
\int d^nx \hspace{1pt}\sqrt{-|G|}\hspace{1pt} e^{-2\phi}
\Bigl(R-{\textstyle{1\over    12}} \hspace{0.5pt} H_{abc}\, H^{abc}
+4 \hspace{0.5pt} \partial_a \phi\hspace{2pt}\partial^a\phi
 \Bigr) \;.
}
By solving the corresponding equations of motion, solutions
have been found which provide the
foundation for many  areas in string theory research.
For instance, Calabi-Yau geometries with vanishing $H$-flux
are the usual starting point for the topological string or for string phenomenology.
Let us note, however,  that in most of these solutions
it is assumed that the string is moving in a Riemannian geometry  supporting
additional $p$-form gauge fields. But, also certain conformal
field theories (CFT) which cannot be identified with such
geometries
provide valid string backgrounds. Typical examples are for instance
asymmetric orbifolds,
but one can also imagine asymmetric CFTs which are not even locally geometric.

To obtain solutions to the field equations in this non-geometric  regime,
T-duality has played an important role. In particular, applying a T-duality transformation to a flat torus with non-vanishing three-form flux $H_{abc}$ leads to a space with so-called geometric flux $f_{ab}{}^c$. A second T-duality results in a background with
non-geometric flux $Q_a{}^{bc}$, where the transition functions between two charts of the manifold have to be extended by T-duality transformations, and hence such spaces are  called T-folds \cite{Dabholkar:2002sy,Hellerman:2002ax,Hull:2004in}.
After formally applying a third T-duality, not along an isometry direction
anymore, one arrives at an  $R$-flux background which does not
admit a clear target-space interpretation.
This chain of T-duality transformations  can be summarized as \cite{Shelton:2005cf}
\eq{
  H_{abc} \;\xleftrightarrow{\;\; T_{c}\;\;}\;
   f_{ab}{}^{c} \;\xleftrightarrow{\;\; T_{b}\;\;}\;
  Q_{a}{}^{bc} \;\xleftrightarrow{\;\; T_{a}\;\;}\;
  R^{abc} \; .
}
For the non-geometric $R$-flux, it has been argued
both from a non-com\-mu\-ta\-tive geometry
\cite{Bouwknegt:2004ap,Mylonas:2012pg,Chatzistavrakidis:2012qj} 
and from a conformal field theory
\cite{Blumenhagen:2010hj,Lust:2010iy,Blumenhagen:2011ph,Condeescu:2012sp} point of view
that a non-associative structure is induced.
However, in contrast to the well-established
non-commutative behavior of open strings \cite{Seiberg:1999vs}, the
generalization of non-commutativity and non-associativity
to the closed string sector  is more difficult,
since in a gravitational theory
the non-commutativity parameter is expected to  be dynamical.
Moreover, a desired deformation quantization is based on the existence
of a \mbox{(quasi-)}symplectic structure,
which is not present in the ordinary description of the closed string.

A  framework to  describe non-geometric $Q$- and $R$-fluxes in a unified way is
provided by generalized geometry \cite{Grana:2008yw,Coimbra:2011nw,Berman:2012vc}
and by  double field theory (DFT) \cite{Siegel:1993th,Hull:2009mi,Hohm:2010jy,Aldazabal:2011nj}.
In the first approach, the concept of Riemannian geometry is extended to a manifold
equipped with
the bundle $TM\oplus TM^*$, whereas in the second
the dimension of the space is doubled by including winding coordinates subject to certain
constraints.
For the latter construction, this results in a manifest $O(D,D)$ invariance of the
action, i.e. the action is invariant under T-duality transformations.
Also, in  double field theory the degrees of freedom are described by  sets of fields,
so-called {\em frames}, which  are related by $O(D,D)$ transformations.
For instance, the hereafter called non-geometric frame contains a metric on the co-tangent bundle,
a dilaton and  a \mbox{(quasi-)}symplectic structure $\fa\beta^{ab}$,
where the latter gives rise to the non-geometric $Q$- and $R$-fluxes.
Since in this frame the $B$-field has been removed,
it is natural to expect that the local diffeomorphism and gauge symmetries
of the string action can be expressed via
a (generalized) differential geometry. This question
has already been approached in an interesting way in
\cite{Andriot:2012wx,Andriot:2012an} (see also \cite{Andriot:2011uh}), however,
the action studied there is not  manifestly invariant under both local symmetries.

The aim of the present paper is to provide details on the construction
of an action which is indeed manifestly bi-invariant under diffeomorphisms and what we
call $\beta$-diffeomorphism. This action for the \emph{non-geometric string} has recently
appeared in the letter \cite{Blumenhagen:2012nk} of the authors and takes
the form
\eq{
  \label{finalaction_intro}
   \hat S={1\over 2\kappa^2} &\int d^nx\, \sqrt{-|\hat g|}\, \bigl|\hat \beta^{-1}\bigr|\, e^{-2\phi}\,
   \Bigl(   \hat R -\tfrac{1}{12} \fa\Theta^{abc}\, \fa\Theta_{abc}
              +4\hspace{1pt} \hat g_{ab}\, D^a\phi D^b \phi\Bigr)\,,
}
where $\fa\beta^{ab}$ is a \mbox{(quasi-)}symplectic structure, $\fa \Theta^{abc}$ denotes the corresponding $R$-flux and the derivative reads $D^a=\fa\beta^{ab} \partial_b$. In \cite{Blumenhagen:2012nk} we have called this theory  {\it symplectic gravity} with a dilaton.
Its action closely resembles the universal part of the
low-energy effective action of string theory, and
the actions \eqref{finalaction_intro} and \eqref{stringaction_intro} are related by a
Seiberg-Witten type redefinition of fields.

\smallskip
In this paper we investigate the underlying mathematical structure and
the properties of the symplectic gravity action \eqref{finalaction_intro}
in the following way:
in section \ref{sec_liealg} we start with a brief introduction to
Lie  algebroids \cite{Halmagyi:2008dr, Blumenhagen:2012pc}, which provide the
mathematical framework for our studies.
In particular,  we outline a differential geometry
calculus giving rise to torsion and curvature tensors
behaving correctly under ordinary diffeomorphisms.
In section \ref{sec_betadiffeos}, we introduce and study $\beta$-diffeomorphisms which
are, besides ordinary diffeomorphisms, the additional symmetry of the symplectic gravity action.
In section \ref{sec_sympgrav}, we  explain the details 
of the differential-geometry construction for $\beta$-diffeomorphisms, formulate the bi-invariant action 
 \eqref{finalaction_intro} and determine the resulting equations of motion.
In section \ref{sec_relst} we discuss the relation between the symplectic gravity action \eqref{finalaction_intro} and string theory,
and derive certain higher-order $\alpha'$-corrections as well as the effective action
of the superstring.
Finally,  in section \ref{sec_eom}  we  study some simple solutions to the field equations
to determine  which types of backgrounds 
are well-described by the symplectic gravity frame.
These examples include approximate solutions with constant $R$-flux
as well as Calabi-Yau geometries in the new frame.
Section \ref{sec_concl} contains our conclusions.


\section{Lie algebroids}
\label{sec_liealg}

We start by giving a brief introduction to Lie algebroids, which can be considered  as a generalization of a Lie algebra by allowing its structure constants to be space-time dependent. Alternatively, a Lie algebroid  can be understood as an extension of the tangent bundle of a manifold to vector bundles, where the latter are equipped with a bracket having similar properties  as the standard Lie bracket. Hence, this approach is suited to apply  constructions known from  differential geometry. 
In physics, Lie algebroids have a plethora of applications, the most relevant for our purposes being the description of non-geometric fluxes (see for example \cite{Halmagyi:2008dr,Halmagyi:2009te,Hull:2004in,Berman:2010is,Blumenhagen:2012pc}).


\subsection{Definition and examples}
\label{sec_lie_defex}

In this section, we introduce the concept of a Lie algebroid and illustrate its properties by two examples.
Let us give the precise definition:
\begin{itemize}

\item[]{\bf Definition:} Let $M$ be a manifold, $E \rightarrow M$ a vector bundle together with a bracket $[\cdot,\cdot]_E : E \times E \rightarrow E$ satisfying the Jacobi identity, and a homomorphism $\rho : E \rightarrow TM$ called the anchor-map. Then $(E,[\cdot,\cdot]_E,\rho)$ is called  \emph{Lie algebroid} if the following Leibniz rule is satisfied
\eq{
\label{Leibniz}
[s_1, f s_2]_E = f \hspace{1pt}[s_1,s_2]_E + \rho(s_1)(f) s_2  \,,
}
for $f\in {\cal C}^{\infty}(M)$ and sections $s_i$ of $E$. For simplicity, if the context is clear we often denote the Lie algebroid just by the total space $E$.
\end{itemize}
Therefore, in a Lie algebroid vector fields and their Lie bracket $[\cdot,\cdot]_L$ are replaced by sections in $E$ and the corresponding bracket. The relation between the different brackets is established by the anchor preserving the algebraic structure
\eq{
	\rho\bigl([s_1,s_2]_E\bigr) = \bigl[\rho(s_1),\rho(s_2)\bigr]_L \;,
}
which can be shown using \eqref{Leibniz} and the Jacobi identity for $[\cdot,\cdot]_E$.

Let us mention two properties of the Lie algebroid $(E,[\cdot,\cdot]_E,\rho)$  which are equivalent to its definition, and which are important for our later constructions.
First, the bracket $[\cdot,\cdot]_E$ on $E$ can be extended to the space of alternating multi-sections $\Gamma(\wedge^{\star}E)$ by defining
\eq{
\gl f, g \gr = 0 \;, \hspace{40pt}
\gl f , s \gr = -\rho(s)\,f \;, \hspace{40pt}
\gl s_1, s_2 \gr = [ s_1, s_2 ]_E \;,
}
for functions $f,g$ and sections $s,s_i$.
For sections of arbitrary degree $a\in \Gamma(\wedge^k E)$, $b\in \Gamma(\wedge^l E)$ and $c\in \Gamma(\wedge^{\star}E)$, the bracket is determined by the relations
\eq{
\label{Gerstenhaber}
  \gl a, b\wedge c \gr &= \gl a, b \gr \wedge c + (-1)^{(k-1)l}\, b\wedge \gl a, c \gr  \;, \\[3pt]
 \gl a,b \gr &= - (-1)^{(k-1)(l-1)}\, \gl b, a \gr \;,
}
which, together with the graded Jacobi identity
\begin{equation}
\gl a, \gl b, c \gr \gr = \gl\gl  a, b \gr , c \gr + (-1)^{(k-1)(l-1)}
\hspace{1pt} \gl b, \gl a, c \gr\gr\; ,
\end{equation}
constitute a so-called Gerstenhaber algebra.
Second, the dual space $\Gamma(\wedge^{\star} E^*)$ is a
graded differential algebra and the differential $d_E$ with respect
to the multiplication $\wedge$ is determined by
\eq{
\label{algebroiddiff}
(d_E \,\omega)(s_0, \dots, s_k) =\hspace{10pt}& \sum_{i=0} ^k \;(-1)^i \rho(s_i)\left( \omega(s_0,\dots,\hat{s}_i,\dots,s_k) \right)  \\
 +&\sum_{i<j} \; (-1)^{i+j} \omega \left([s_i, s_j]_E,s_0,\dots,\hat{s}_i,\dots,\hat{s}_j,\dots,s_k \right)
 \;,
}
where $\omega \in \Gamma(\wedge^k E^*)$, $\{s_i\}\in \Gamma(E)$ and where the hat stands for deleting the corresponding entry.

Furthermore, there are two standard examples for Lie algebroids which will be used in later sections in this paper. We discuss them in turn.
\begin{itemize}

\item First, consider $ E=(TM,[\cdot,\cdot]_{L},\rho= \textrm{id})$ where the anchor is
the identity map and the bracket is given by the usual Lie bracket
$[X,Y]_{L}$ of vector fields. The extension to multi-vector fields in
$\Gamma(\wedge^{\star}TM)$ is given by
the relations \eqref{Gerstenhaber}, which results in the so-called Schouten--Nijenhuis  bracket $[\cdot,\cdot]_{SN}$. 
The differential on the dual space $\Gamma(\wedge^{\star}T^* M)$ is the standard de Rham differential.

\item For the second example, let $(M,\beta)$ be a Poisson manifold
with Poisson tensor $\beta =  \frac{1}{2}\,\beta^{ab}e_a \wedge e_b$.
Note that if $\beta$ is a proper Poisson tensor, it follows that $\Theta=\frac{1}{2}\,[\beta,\beta]_{SN}=0$.
The Lie algebroid is  given by
$E^*=(T^* M,[\cdot,\cdot ]_K,\rho =  \beta^\sharp)$, where
the anchor $\beta^{\sharp}$ is defined as
\eq{ \label{anchor}
\beta^{\sharp} (e^a) = \beta^{am}e_m \;.
}
for $\{e^a\}$ a basis of one-forms.
The bracket on $T^*M$ is the \emph{Koszul bracket}, which for one-forms is defined as
\eq{
\label{koszul}
	[\xi,\eta]_K = L_{\beta^\sharp(\xi)}\eta
 -\iota_{\beta^\sharp(\eta)}\,d\xi \; ,
}
where the Lie derivative on forms is given by $L_{X} = \iota_X \circ d + d \circ \iota_X$ with $d$ the de Rham differential.
The associated bracket for forms with arbitrary degree is again determined by \eqref{Gerstenhaber} and is called the \emph{Koszul--Schouten bracket}. 
The corresponding differential on the dual space $\Gamma(\wedge^{\star}TM)$ is given in terms of the Schouten--Nijenhuis bracket as
\eq{
  \label{betad}
  d_{\beta} = [\beta,\cdot\,]_{SN} \; .
}
\end{itemize}


\subsection{Generalizing constructions of differential geometry}
\label{sec_gendiffg}

We are now going to generalize notions of differential geometry such as  the Lie and covariant derivative to Lie algebroids. The standard constructions  in this setting can be found for instance in \cite{2008arXiv0806.3522B} (see also \cite{2000math......1129L,Gualtieri:2007bq}), but here we only  recall the most important ones to set our conventions and to motivate the calculus to be formulated in the next section. 
However, let us note that here we work with proper Lie algebroids for which the Jacobi identity is satisfied. Especially in section~\ref{sec_betadiffeos}, we also employ quasi-Lie algebroids where the Jacobi identity is violated and where some of the formulas presented here are not valid. 
We will come back to this point below.


\vspace*{-3pt}
\subsubsection*{Lie derivative}
\vspace*{-3pt}

We begin with the generalization of the Lie derivative. For a section $s$ of $E$ we define its action on functions $f$ by
\eq{
\label{Lie}
{\cal L}_s f := s(f) := \rho(s)(f) \;,
}
which in the trivial example of $TM$  coincides with the original Lie derivative, that is $\mathcal L_{e_a} f= \partial_a f$. For our second example of $T^*M$, formula \eqref{Lie} allows us to define derivatives in the direction of a one-form. In particular,
for $e^a$ we have
\eq{\label{D}
\Lie_{e^a} f = \beta^{\sharp}(e^a)(f) = \beta^{ab}\partial_b f =: D^a f \;,
}
where we introduced $D^a=\beta^{ab}\partial_b$.
Note that \eqref{Lie} is compatible with the Lie bracket on $E$ because of the following relation for a function $f$
\eq{\label{consistency}
\bigl [\Lie_{s_1},\Lie_{s_2}\bigr] \hspace{1pt}f = \Lie_{[s_1, s_2]_E} f  \;.
}
The Lie derivative acting on sections of $E$ is defined using the bracket on the total space $E$, while for sections of the dual $E^*$ the Cartan formula and the associated differential $d_E$ on $E^*$ are employed. More precisely, for sections $s,s_i$ of $E$ and $\alpha$ of $E^*$ we have
\eq{ \label{Lie2}
\Lie_{s_1} s_2 &= [s_1, s_2]_E \; , \hspace{40pt}
\Lie_s \alpha = \iota_s \circ d_E \alpha + d_E \circ \iota_s \alpha \;,
}
where the insertion map $\iota$ is defined in the standard way, that is for a basis $\{s_a\}$ of $E$ and dual basis $\{s^a\}$ of $E^*$ we have $\iota_{s_a} s^b = \delta^b_a$.
The extension of \eqref{Lie2} to multi-sections is given by using the product rule.

With the definitions \eqref{Lie} and \eqref{Lie2} it is now easy to prove the following properties of the Lie derivative for a Lie algebroid. In particular, employing the Jacobi identity  we have 
\eq{\label{Lierel}
\Lie_s \circ d_E = d_E \circ \Lie_s \;, \hspace{40pt}
\iota_{[s_1,s_2]_E} = \Lie_{s_1} \circ \iota_{s_2} - \iota_{s_2} \circ \Lie_{s_1}  \;, 
}
when acting on elements in $\Gamma(\wedge^{\star} E^*)$, 
and for elements both in $\Gamma(\wedge^{\star} E)$ and $\Gamma(\wedge^{\star} E^*)$  we find
\eq{
  \label{lierel2}
\bigl[\Lie_{s_1},\Lie_{s_2} \bigr] = \Lie_{[s_1,s_2]_E} \;.
}


\subsubsection*{Covariant derivative}

We can now proceed and generalize the notion of  covariant differentiation to a Lie algebroid $E$ \cite{2008arXiv0806.3522B}.
\begin{itemize}

\item[]{\bf Definition:} A \emph{covariant derivative} on $E$ is a bilinear map $\nabla: \Gamma(E)\times \Gamma(E) \rightarrow \Gamma(E)$ which has the properties
\eq{\label{covder}
\nabla_{fs_1} s_2 =  f\hspace{1pt}\nabla_{s_1} s_2  \,, \hspace{40pt}
\nabla_{s_1} fs_2 =  \rho(s_1)(f) s_2 + f\hspace{1pt}\nabla_{s_1} s_2 \,.
}

\end{itemize}
Following this definition, it is possible to obtain curvature and torsion operators. They are given by formulas similar to the standard case on the tangent bundle 
\eq{ \label{curv}
R(s_a,s_b)s_c &=\; \nabla_{s_a}\nabla_{s_b} s_c - \nabla_{s_b}\nabla_{s_a} s_c
- \nabla_{[s_a,s_b]_{E}}\, s_c  \;, \\[3pt]
T(s_a,s_b) &= \; \nabla_{s_a}s_b - \nabla_{s_b}s_a - [s_a,s_b]_{E}  \,.
}
To see that these expressions are tensors with respect to standard diffeomorphisms it suffices to check that they are ${\cal C}^{\infty}$-linear in every argument. The reason is  that for a general ${\cal C}^{\infty}$ multi-linear map $ A : \Gamma\bigl( (\otimes^r TM) \otimes (\otimes^s T^*M)\bigr) \rightarrow {\cal C}^{\infty}(M)$ and coordinates $x^\mu, y^{\mu '}$ we have
\eq{\label{tensor}
A^{\mu_1 \dots \mu_r}{}_{\nu_1 \dots \nu_s} &= A(dx^{\mu_1},\dots,dx^{\mu_r},\partial_{\nu_1},\dots, \partial_{\nu_s}) \\
&= A\left(\tfrac{\partial x^{\mu_1}}{\partial y^{\mu_1 '}} dy^{\mu_1 '},\dots,\tfrac{\partial x^{\mu_r}}{\partial y^{\mu_r '}}dy^{\mu_r '},\tfrac{\partial y^{\nu_1 '}}{\partial x^{\nu_1}} \partial_{\nu_1 '},\dots,\tfrac{\partial y^{\nu_s '}}{\partial x^{\nu_s}} \partial_{\nu_s '}\right) \\
&=\tfrac{\partial x^{\mu_1}}{\partial y^{\mu_1 '}} \cdots \tfrac{\partial x^{\mu_r}}{\partial y^{\mu_r '}}\tfrac{\partial y^{\nu_1 '}}{\partial x^{\nu_1}} \cdots \tfrac{\partial y^{\nu_s '}}{\partial x^{\nu_1}} A^{\mu_1' \dots \mu_r'}{}_{\nu_1' \dots \nu_s'} \;.
}
The proof of  ${\cal C}^{\infty}$-linearity for both expressions in \eqref{curv} is now a straightforward calculation using the definition \eqref{covder} and the Leibniz rule \eqref{Leibniz}.


\subsubsection*{Metric}

Finally, a metric on a Lie algebroid $E$ is an element in $\Gamma(E^*\otimes_{\textrm{sym}}E^*)$ which gives rise to a scalar product for sections in $E$. The latter will be denoted by
\eq{
\label{metric}
\langle s_a, s_b \rangle = g_{ab} \;.
}
Employing this definition, we  obtain a unique connection $\mathring\nabla$, generalizing the standard Levi-Civita connection, if we demand 
\begin{itemize}
\item vanishing torsion: \hspace{30pt}$\mathring\nabla_{s_1} s_2 - \mathring\nabla_{s_2} s _1 = [s_1,s_2]_E$,
\item metricity: \hspace{70.5pt}$\rho(s_1)\langle s_2,s_3 \rangle = \langle \mathring\nabla_{s_1}s_2, s_3 \rangle + \langle s_1, \mathring\nabla_{s_2} s_3 \rangle $.
\end{itemize}
The connection $\mathring\nabla$ is characterized by the Koszul formula,
where the proof again follows along the lines of standard differential geometry
\eq{\label{Koszulformula}
2\hspace{1pt}\bigl\langle \mathring\nabla_{s_1} s_2,s_3\bigr\rangle = & \;s_1\bigl(\langle s_2,s_3 \rangle \bigr)
+ s_2\bigl(\langle s_3,s_1 \rangle \bigr) - s_3\bigl(\langle s_1 ,s_2 \rangle \bigr)  \\[3pt]
&- \langle s_1 ,[s_2,s_3]_E\rangle + \langle s_2,[s_3,s_1]_E\rangle + \langle s_3,[s_1,s_2]_E\rangle  \;.
}


\subsubsection*{Summary}

As we have reviewed in this section, a Lie algebroid admits constructions similar to standard differential geometry on the tangent bundle. This is plausible since the latter is a special case of a Lie algebroid with trivial anchor. Furthermore, objects such as the curvature and torsion tensor can be defined, which indeed have desirable properties such as multi-linearity. In particular, from \eqref{curv} we obtain an analogue of  the Ricci scalar allowing us to formulate Einstein gravity in this framework, and in section \ref{sec_sympgrav} we employ  a particular Lie algebroid to construct torsion and curvature tensors appropriate to our study of non-geometric fluxes.


\section{$\beta$-diffeomorphisms}
\label{sec_betadiffeos}

Our aim in this section is to develop a covariant tensor calculus on $T^*M$, which admits the usual behavior under diffeomorphisms but also includes a proper analogue of gauge transformations. The reason for implementing the latter stems from translating the geometric objects of interest on the tangent bundle, that is the metric and the Kalb-Ramond field, to the co-tangent bundle. In the following, we will motivate a new type of diffeomorphisms and introduce the appropriate notion of a  covariant tensor.
However, let us emphasize that in contrast to section~\ref{sec_liealg}, here we will mostly work with a quasi-Lie algebroid for which the Jacobi identity is not satisfied.


\subsection{From gauge transformations to $\beta$-diffeomorphisms}
\label{subsec:gtob}

Developing a framework for describing T-dual configurations  in string theory
necessitates the implementation of the underlying symmetries. Usually,
string-theoretical geometries are characterized by a metric $G$ and a
Kalb-Ramond two-form $B$ on the target-space manifold $M$. Both behave
covariantly under  diffeomorphisms, but additionally, $B$ is considered to be
an abelian two-form gauge field. Thus,  the theory has to be invariant under the gauge transformations
\eq{\label{gtB}
	B\mapsto B+d\hspace{1pt}\xi \;.
}


\subsubsection*{Translation from the tangent to the co-tangent bundle}

Let now $\{e_a\}$ be a holonomic frame for the vector fields in $\Gamma(TM)$ and $\{e^a\}$ be the dual frame. Assuming $B=\frac{1}{2}B_{ab}\hspace{1pt}e^a\wedge e^b$ to be invertible, we introduce the \emph{quasi-Poisson} structure
\eq{\label{qps}
	\beta = B^{-1} = \tfrac{1}{2}\,\beta^{ab}\,e_a\wedge e_b \,.
}
We require this only to be a quasi-Poisson structure because we aim to describe a theory with non-vanishing $R$-flux\hspace{2pt}\footnote{Here and in the following, (anti-)symmetrization of indices is defined with a factor of $1/n!$.}
\eq{
  \label{flux_78}
	\Theta^{abc} = \tfrac{1}{2} \hspace{1pt} \bigl([\beta,\beta]_{SN} \bigr)^{abc}
	= 3\,\beta^{[a|m}\partial_m\beta^{|bc]} \, ,
}
where $[\cdot,\cdot]_{SN}$ is the Schouten--Nijenhuis  bracket introduced in section \ref{sec_lie_defex}. Indeed, the Jacobi identity of the induced Poisson bracket $\{f,g\}=\beta(df,dg)$  is not satisfied but evaluates to
\eq{
  \label{jacobi_03}
	\mathrm{Jac}(f,g,h) &= \{f,\{g,h\}\} + \{h,\{f,g\}\} +\{g,\{h,f\}\} \\
	&= \Theta^{abc}\, (\partial_af)\, (\partial_bg)\, (\partial_ch) \;,
}
which justifies the name.
The quasi-Poisson structure $\beta$ furthermore introduces an anchor map which relates the tangent and co-tangent bundle
\eq{
	\beta^\sharp:T^*M\to TM\,, \hspace{40pt} \eta\mapsto\beta^\sharp\eta
	= \beta^{ma}\,\eta_m\,e_a\;,
}
and if we consider the anchor to be invertible it can be used to translate between geometric objects. In particular, in our construction $\beta$  replaces the Kalb-Ramond field $B$ and we  obtain a metric $\fa g$ on $T^*M$ by \emph{anchoring} $G$ on $TM$, that is
\eq{\label{anchG}
  \fa g =  \bigl( \otimes^2\beta^\sharp \bigr) (G)
  = G_{mn}\,\beta^\sharp e^m\otimes \beta^\sharp e^n
  = (\beta^{am}\,\beta^{bn}\,G_{mn})\,e_a\otimes e_b \, .
}
In this way, we have replaced
\eq{\label{redef}
	B_{ab} &\;\to\; \beta^{ab} = (B^{-1})^{ab} \;, \\
	G_{ab} &\;\to\; \fa g^{ab} = \beta^{am}\,\beta^{bn}\,G_{mn} \;,
}
which are covariant tensors since $G$ and $B$ are covariant. The implications of this translation will be studied in more detail in section~\ref{sec_relst}.


\subsubsection*{$\beta$-diffeomorphisms}

Because the metric $\fa g^{ab}$ on the co-tangent bundle is expressed in terms of $G$ and the Kalb-Ramond field $B$, it changes under the gauge transformations \eqref{gtB}. Recalling then 
\eq{
  \delta_{\xi}^{\mathrm{gauge}}\hspace{1pt}B_{ab}=\partial_a\hspace{1pt}\xi_b-\partial_b\hspace{1pt}\xi_a \;,
}
and using $\delta_{\xi}^{\mathrm{gauge}}B_{ab}
	= - B_{am}\bigl(\delta_{\xi}^{\mathrm{gauge}}\beta^{mn}\bigr)B_{nb}$,
we obtain from \eqref{redef} that
\eq{\label{gtbg}
	\delta_{\xi}^{\mathrm{gauge}}\hspace{1pt}\beta^{ab} &= \beta^{am}\beta^{bn}
		\bigl(\partial_m\xi_n-\partial_n\xi_m\bigr) \;, \\
	\delta_{\xi}^{\mathrm{gauge}}\hspace{1pt}\fa g^{ab} &= 2\hspace{1pt}\fa g^{(a|m}\beta^{|b)n}
		\bigl(\partial_m\xi_n-\partial_n\xi_m\bigr) \;.
}
Let us furthermore recall from section~\ref{sec_lie_defex} that the ordinary Lie derivative $L_X$ acts on vector  fields through the Lie bracket, and that via the relations \eqref{Lie2} we can construct a derivative  $\fa{\mathcal L}_{ \xi}$ based on the Koszul bracket \eqref{koszul}. In particular, for \raisebox{0pt}[\ht\strutbox]{$\fa{\mathcal L}_{ \xi}$} acting on  a one-form $\eta$ and a vector field $X$ we have
\eq{
\label{koszul_lie_04}
\fa{\mathcal L}_{\xi} \eta = \bigl[\xi, \eta\bigr]_K \; , \hspace{40pt}
\fa{\mathcal L}_{\xi} X = \iota_{\xi} \circ d_{\beta} X + d_{\beta} \circ \iota_{\xi} X \;,
}
where the differential $d_{\beta}$ is defined by \eqref{algebroiddiff} (cf.~\eqref{betad}).
Note  that due to the non-vanishing $R$-flux  \eqref{flux_78}, the derivative $\fa{\mathcal L}_{\xi}$ does not satisfy the relations \eqref{Lierel} and \eqref{lierel2}, and hence is not a proper Lie derivative.
Nevertheless, employing \eqref{koszul_lie_04}
we can rewrite equations \eqref{gtbg} as
\begin{align}
\label{gtg}
\begin{split}
  \delta_{\xi}^{\mathrm{gauge}}\hspace{1pt}\fa g^{ab}
  &= (L_{\beta^\sharp\xi}\hspace{1pt}\fa g)^{ab}
  - (\fa{\mathcal L}_{\xi}\hspace{1pt}\fa g)^{ab} \\
  &= (L_{\beta^\sharp\xi}\hspace{1pt}\fa g)^{ab} - \fa\delta_{\xi}\hspace{1pt}\fa g^{ab} \;,
\end{split}
\\[3pt]
\label{gtb}
\begin{split}
  \delta_{\xi}^{\mathrm{gauge}}\hspace{1pt}\beta^{ab}
  &= (L_{\beta^\sharp\xi}\hspace{1pt}\beta)^{ab}
  - \bigl[ (\fa{\mathcal L}_{\xi}\beta)^{ab}
  + \beta^{am}\beta^{bn}\bigl(\partial_m\xi_n-\partial_n\xi_m\bigr)\bigr]\\
  &= (L_{\beta^\sharp\xi}\hspace{1pt}\beta)^{ab} - \fa\delta_{\xi}\hspace{1pt}\beta^{ab} \;,
\end{split}
\end{align}
where we have introduced 
\eq{
\label{gtgb}
   \fa\delta_{\xi}\hspace{1pt}\fa g^{ab} &= (\fa{\mathcal L}_{\xi}\hspace{1pt}\fa g)^{ab} \;,\\
   \fa\delta_{\xi}\hspace{1pt}\beta^{ab}  &=
   (\fa{\mathcal L}_{\xi}\beta)^{ab}
  + \beta^{am}\beta^{bn}\bigl(\partial_m\xi_n-\partial_n\xi_m\bigr) \;.
}
Thus, after the replacement \eqref{redef} we can split  gauge transformations  into a subgroup of usual diffeomorphisms and a new transformations, infinitesimally denoted by $\fa\delta_{\xi}$, 
which will be called \emph{$\beta$-diffeomorphisms} and 
to which the remainder of this section is devoted to.


\subsection{Interlude: the partial derivative}

As will become clear below, the proper analogue of the  partial derivative in the present context is the derivative \eqref{D}, whose action on a function $f$ we recall for convenience
\eq{
	D f = (D^af) \, e_a = \beta^{am} (\partial_m f) \, e_a\;.
}
This derivative  is the differential associated to the Koszul bracket through \eqref{algebroiddiff}, and  will be covariantized in section~\ref{sec_sympgrav}.

If we consider now again a holonomic frame $\{e_a\}$ of $\Gamma(TM)$ and its dual $\{e^a\}$, the partial derivative can be considered as the action of a basis vector field on functions, i.e. $e_a(f)=\partial_af$. Analogously, we can act with anchored forms to obtain $D^a$
\eq{
	(\beta^\sharp e^a)f=D^af \, .
}
These vector fields satisfy an algebra (which already appeared in \cite{Grana:2008yw} and \cite{Blumenhagen:2012pc}) of the form
\eq{\label{der_alg}
	[e_a,e_b]_L &= 0 \;, \\
	[e_a,\beta^\sharp e^b]_L &= Q_a{}^{bm}\,e_m  \;, \\
	[\beta^\sharp e^a,\beta^\sharp e^b]_L &= \Theta^{abm}\,e_m
		+ Q_m{}^{ab}\,(\beta^\sharp e^m)
	\;,
}
where $\Theta$ is the $R$-flux given in \eqref{flux_78}
and the $Q$-flux is defined as
\eq{
	Q_c{}^{ab} = \bigl([e^a,e^b]_K\bigr)_c = \partial_c \hspace{1pt}\beta^{ab} \;.
}
Note that from the last equation in \eqref{der_alg} we infer that $\beta^\sharp$ fails to be an algebra-homomorphism for the Koszul bracket if $\Theta\neq0$. Thus, the $R$-flux can be interpreted as the corresponding defect.
Finally, the Jacobi identities  associated to \eqref{der_alg}, also referred to as Bianchi identities in the following, will be of importance for the rest of the paper and
read \cite{Blumenhagen:2012ma,Blumenhagen:2012pc}
\eq{\label{bianchi_plain}
	0 &= 3\,D^{[a}Q_d{}^{bc]} + 3\,Q_d{}^{[a|m}\,Q_m{}^{|bc]} - \partial_d\Theta^{abc} \;, \\
	0 &= 2\,D^{[a}\Theta^{bcd]}-3\,\Theta^{[ab|m}\,Q_m{}^{|cd]} \;.
}


\subsection{$\beta$-tensors}
\label{sec:betaten}

In section~\ref{subsec:gtob} we have seen how diffeomorphisms and gauge transformations can be translated from the tangent to the co-tangent bundle, and how a re-interpretation of the gauge transformations leads to a new type of diffeomorphisms. For our purpose of constructing a gravitational theory, we require the new metric $\fa g$ in \eqref{redef} to transform properly also with respect to the new transformation $\fa\delta_{\xi}$. The expressions in \eqref{gtgb} then suggests that the latter should be characterized by the derivative \raisebox{0pt}[\ht\strutbox]{$\fa{\mathcal L}_{\xi}$}, which we will use as a guiding principle in the following.


\subsubsection*{Definition and examples}

Before giving the definition of a $\beta$-tensor, let us first recall the situation in the standard case. Here, the transformation properties of a tensor can be characterized by the associated group, and for infinitesimal transformations by the algebra.
More concretely, an $(r,s)$-tensor field $T$ is a section in $(\otimes^rTM)\otimes(\otimes^s T^*M)$, implying that it is a multi-linear form. A tensor field is covariant since it is invariant under diffeomorphisms, and because the associated Lie algebra is $\Gamma(TM)$, covariance infinitesimally translates  to 
\eq{
   \delta_X \hspace{1pt} T^{a_1\dots a_r}{}_{b_1\dots b_s}
   =(L_X\hspace{1pt} T)^{a_1\dots a_r}{}_{b_1\dots b_s}\;,
}
with $L_X$ the usual Lie derivative in the direction of a vector field $X$. 
In the following, we  adopt the description in  terms of the algebra 
and define a $\beta$-tensor via the derivative \eqref{koszul_lie_04}. More concretely,
\begin{itemize}

\item[]{\bf Definition:} A tensor $T\in\Gamma\bigl( \hspace{1pt}(\otimes^rTM)\otimes(\otimes^s T^*M) \bigr)$ is called a \emph{$\beta$-tensor} if for  a one-form $\xi$ it behaves as
\eq{\label{btensor}
	\fa\delta_{\xi}\hspace{1pt}T^{a_1\dots a_r}{}_{b_1\dots b_s}
	= \bigl(\fa{\mathcal{L}}_{\xi} \hspace{1pt}T \bigr)^{a_1\dots a_r}{}_{b_1\dots b_s}  \;,
}
where $\fa{\mathcal{L}}_{\xi}$ is the derivative defined by the Koszul bracket \eqref{koszul_lie_04} which takes the form
\eq{
 \bigl(\fa{\mathcal{L}}_{\xi} \hspace{1pt}T \bigr)^{a_1\dots a_r}{}_{b_1\dots b_s}
  =  & \hspace{15pt} \xi_m\,D^m \hspace{1pt}T^{a_1\dots a_r}{}_{b_1\dots b_s} \\
&- \sum_{i=1}^s \bigl(D^m\xi_{b_i}+\xi_n\,Q_{b_i}{}^{mn}\bigr)\hspace{1pt}
T^{a_1\dots a_r}{}_{b_1\dots b_{i-1}\, m\, b_{i+1}\dots b_s} \\
&+ \sum_{i=1}^r \bigl(D^{a_i}\xi_m+\xi_n\,Q_m{}^{a_i n} \bigr) \hspace{1pt}
T^{a_1\dots a_{i-1}\,m\,a_{i+2}\dots a_r}{}_{b_1\dots b_s} \,.
}

\end{itemize}

For  constructing a gravitational theory incorporating the $R$-flux, we require $\Theta^{abc}$ as well as the metric $\fa g^{ab}$ to be $\beta$-tensors. Moreover, as $D^a$ is the analogue of the usual partial derivative on $T^*M$, we also impose that $D^a f$ should be a $\beta$-tensor
if $f$ is a $\beta$-scalar.
These requirements can be used to determine the transformation behavior of $\beta$. In particular, assuming that $[\fa\delta_{\xi},\partial_a]=0$ and employing \eqref{der_alg}, we obtain
\eq{
	\fa\delta_{\xi} \bigl(D^af \bigr)
	= \bigl(\fa{\mathcal L}_{\xi}\hspace{1pt}Df\bigr)^a
	+ \bigl(\fa\delta_{\xi}\hspace{1pt} \beta^{ab}-\Theta^{abm}\xi_m\bigr) 
	\hspace{1pt}\partial_bf \, .
}
With the above restrictions it then follows that $\beta$ cannot be a $\beta$-tensor itself but has to transform as
\eq{\label{tbeta}
\fa\delta_{\xi}\hspace{1pt}\beta^{ab} 
&= \Theta^{abm}\,\xi_m  \\
&=\fa \Lie_{\xi}\hspace{1pt} \beta +\beta^{am}\beta^{bn}\bigl(\partial_m\xi_n-\partial_n\xi_m\bigr)
\,,
}
which is consistent with \eqref{gtgb}.
Finally, given the transformation  of $\beta$ and using the Bianchi identity \eqref{bianchi_plain}, we can show that also the $R$-flux $\Theta$ behaves as  a $\beta$-tensor, that is 
\eq{\label{tR}
	\fa\delta_{\xi}\Theta^{abc} &= (\fa{\mathcal L}_{\xi}\Theta)^{abc}
		+ 2\,\xi_d\left(2D^{[a}\Theta^{bcd]}
		-3\Theta^{[ab|m|}\,Q_m{}^{cd]}\right) \\
	&= (\fa{\mathcal L}_{\xi}\Theta)^{abc} \,.
}


\subsubsection*{Algebra of transformations}

As we aim to describe a theory admitting a non-vanishing $R$-flux, the Koszul bracket \eqref{koszul} does not satisfy the Jacobi identity and the anchor is not a homomorphism. The defect 
to both of these properties is proportional to the $R$-flux, which for the Jacobi identity can be seen 
from
\eq{\label{jacK}
	\mathrm{Jac}_K(\eta,\chi,\zeta)
	&= \bigl[\eta,[\chi,\zeta]_K\bigr]_K +\bigl[\zeta,[\eta,\chi]_K\bigr]_K +\bigl[\chi,[\zeta,\eta]_K\bigr]_K \\
	&= \bigl[\Lie_\eta,\Lie_\chi\bigr]\hspace{1pt}\zeta-\Lie_{[\eta,\chi]_K}\zeta\\
	&= d\bigl(\Theta(\eta,\chi,\zeta)\bigr) + \iota_{(\iota_\zeta\iota_\chi\Theta)}d\eta
		+ \iota_{(\iota_\eta\iota_\zeta\Theta)}d\chi
		+ \iota_{(\iota_\chi\iota_\eta\Theta)}d\zeta \;,
}
where we employed the first Bianchi identity in \eqref{bianchi_plain}.
Similarly, the derivative $\fa{\mathcal L}_{ \xi}$ does not commute with $D$ when acting on functions $f$, which can be computed using \eqref{der_alg} as
\eq{
  \bigl[\fa{\mathcal L}_{\xi},D \bigr] f = -\Theta^{amn}\,\xi_m\,(\partial_n f)\,e_a \, .
}
This hints towards an algebra of infinitesimal transformations which is not closed. Indeed, for a $\beta$-tensor $\eta$ we find
\eq{\label{lieilldef}
	\fa\delta_{\xi_2}\bigl(\fa\delta_{\xi_1}\eta_a\bigr)
	= \bigl(\fa{\mathcal L}_{\xi_2}\fa{\mathcal L}_{\xi_1}\eta\bigr)_a
		+ \xi_{(1)m}\,\eta_n\,\Theta^{mnk}\bigl( \partial_a\xi_{(2)k}
		-\partial_k \xi_{(2)a} \bigr) \, ,
}
where we employed \eqref{bianchi_plain}.
That is, the variation of a $\beta$-tensor is not a $\beta$-tensor but transforms anomalously.
More generally, for vector fields $X_1$, $X_2$ and one-forms $\xi_1$, $\xi_2$ we can deduce 
\eq{\label{trafo_alg}
	\bigl[\delta_{X_1},\delta_{X_2}\bigr] &= \delta_{[X_1,X_2]_L} \;, \\
	\bigl[\fa\delta_{\xi_1},\delta_{X_1}\bigr] &= \delta_{(\fa{\mathcal L}_{\xi_1}X_1)} \;,\\
	\bigl[\fa\delta_{\xi_1},\fa\delta_{\xi_2}\bigr] &= \fa\delta_{[\xi_1,\xi_2]_K}
		+ \delta_{(\iota_{\xi_1}\iota_{\xi_2}\Theta)} \; .
}
The defect of the algebra of $\beta$-transformations to close can be traced back to the failure of the Jacobi identity \eqref{jacK} for the Koszul bracket for non-vanishing $\Theta$. However, this defect can be written as a diffeomorphism, which means that the algebra closes considering $\beta$-diffeomorphisms along with usual diffeomorphisms. This is of course expected from the translation of gauge transformations as can be seen in \eqref{gtg}.


\subsubsection*{Remarks and summary}

Let us close this section with two remarks and a short summary.
\begin{itemize}

\item The term {\em $\beta$-diffeomorphism} has been chosen to emphasize the similarity between usual diffeomorphisms and the new transformations on the co-tangent bundle. In particular, the latter are characterized by a derivative based on the Koszul bracket as
\eq{
	\fa{\mathcal L}_{\xi}\hspace{1pt}\eta = L_{\beta^\sharp\xi}\hspace{1pt}\eta
		- \iota_{\beta^\sharp\eta}d\xi \;, \hspace{40pt}
	\fa{\mathcal L}_{\xi}X = L_{\beta^\sharp\xi}X
		+ \beta^\sharp\bigl(\iota_{X}d\xi\bigr) \;,
}
for  $\eta$ a one-form and $X$ a vector field. We observe that usual
diffeomorphisms are only a subgroup of the new transformations, whose algebra
is generated by vector fields of the form $\beta^\sharp\xi$. The remaining
part is given by the remnant of the original gauge transformations. Therefore,
in contrast to equation \eqref{tensor} for standard diffeomorphisms, we cannot give
an analogous  integrated version of  $\beta$-diffeomorphisms.

\item The anchoring procedure we used in \eqref{anchG} to obtain the new metric $\fa g^{ab}$ provides a strong device of translating usual tensors on $TM$ to $\beta$-tensors on $T^*M$. This allows us to derive  the $T^*M$-analogue of usual geometric objects, as we will see in section~\ref{sec_relst}.

\end{itemize}
To summarize our discussion in this section so far, we have introduced $\beta$-trans\-for\-ma\-tions as the co-tangent bundle analogue of gauge transformations, and we have described them  infinitesimally in terms of the Koszul bracket. We furthermore observed that $\beta$ itself does not transform as a $\beta$-tensor in order for the $R$-flux and derivatives of scalars to be proper $\beta$-tensors. In section~\ref{sec_sympgrav}, we will develop a differential geometry calculus incorporating this new symmetry together with diffeomorphisms.


\subsection{The Courant algebroid perspective}

Before closing this section, let us discuss the Courant algebroid which 
provides an interesting link between our constructions and  generalized geometry, but which 
will not be of relevance for the rest of this paper.
More concretely, in equation \eqref{trafo_alg} we have seen that the algebra of $\beta$-diffeomorphisms does not close by itself. However, as we will illustrate now, this issue can be resolved by introducing a Courant algebroid structure  \cite{Roytenberg:01,Halmagyi:2009te,Blumenhagen:2012pc} with a bracket on the generalized tangent bundle $TM\oplus T^*M$.

We first introduce the \emph{Dorfman bracket}\hspace{2pt}\footnote{Let us mention that in \cite{Blumenhagen:2012pc} we have worked with the \emph{Courant bracket}, which is  the symmetrized version of the Dorfman bracket.}   $\cdot\,\bullet\,\cdot$, which in the case of vanishing $H$-flux is determined by the following relations for 
 vector fields $X$, $Y$ and one-forms $\eta$, $\chi$
\eq{\label{dorfman}
\arraycolsep1.5pt
\renewcommand{\arraystretch}{1.2}
\begin{array}{rcll}
	X &\bullet& Y &= [X,Y]_L \;, \\
	X&\bullet& \eta &= \iota_X\circ d\eta + d\circ\iota_X\eta -\iota_\eta\circ d_\beta X \;, \\
	\eta&\bullet& X &= \iota_\eta\circ d_\beta X + d_\beta\circ\iota_\eta X
		- \iota_X\circ d \eta \;, \\
	\eta&\bullet&\chi &= [\eta,\chi]_K
		+ \iota_\chi\iota_\eta\Theta \;,
\end{array}		
}
where $d_\beta$ is the differential defined in \eqref{betad}. This bracket
satisfies the Jacobi identity, and $\mathrm{id}_{TM}+\beta^\sharp$ is an
algebra homomorphism which serves as the anchor \cite{Blumenhagen:2012pc}. Therefore, the corresponding Dorfman-Lie derivative\hspace{2pt}\footnote{For more details on the construction of a Lie and covariant derivative using the Dorfman bracket see for instance  
\cite{2004math......7399V,Ellwood:2006ya,Gualtieri:2007bq,Grana:2008yw,Hull:2009zb,Berman:2012vc}.}
\eq{
  \Lie^D_A B = A\bullet B \hspace{40pt}{\rm for} \quad A,B\in\Gamma(TM\oplus T^*M)
}
satisfies
\eq{\label{jacD}
	\bigl[\Lie^D_A,\Lie^D_B\bigr]C = \Lie^D_{A\bullet B} \hspace{1pt}C \, .
}
In terms of the usual Lie derivative $L$ and the derivative $\fa{\mathcal L}$ based on the Koszul bracket, the defining relations \eqref{dorfman} for the Dorfman bracket can be written in the following way
\eq{\label{LieD}
\arraycolsep1.5pt
\renewcommand{\arraystretch}{1.3}
\begin{array}{ll}
	\Lie^D_X \hspace{1pt}Y &= L_X Y  \;, \\
	\Lie^D_X \hspace{1pt}\eta &= L_X\eta -\iota_\eta\circ d_\beta X\;, \\
	\Lie^D_\eta X &= \fa{\mathcal L}_\eta X - \iota_X\circ d \eta \;,\\
	\Lie^D_\eta\chi &= \fa{\mathcal L}_\eta \chi + \iota_\chi\iota_\eta\Theta \;.
\end{array}	
}
Note that the the first term on the right-hand side in each line is type-preserving, and that the additional  terms are necessary for the Jacobi identity to be satisfied.
However, ignoring the latter ``off-diagonal'' terms we  see that infinitesimal diffeomorphisms are characterized by the first two lines in \eqref{LieD} while infinitesimal $\beta$-diffeomorphisms are given by the last two. Thus, defining ($\beta$-)diffeomorphisms by the Dorfman bracket modulo off-diagonal terms would lead to a closure of the algebra of infinitesimal transformations, since the  Jacobi identity for the Dorfman bracket is satisfied.
For our purpose of constructing an action expressed in terms of quantities on the co-tangent bundle resulting in a $\beta$-scalar, the off-diagonal terms are not important and so we can work with the derivative $\fa{\mathcal L}_{\xi}$.


\section{Bi-invariant geometry and symplectic gravity}
\label{sec_sympgrav}

In this section, we introduce a differential geometry  for the co-tangent
bundle, providing  the geometric notions and objects  consistent with
diffeomorphisms and $\beta$-diffeomorphisms. To this end, we introduce a
suitable Lie algebroid and derive in detail the form of the connection,
torsion and curvature along the lines of section~\ref{sec_gendiffg}. This will
allow us to construct an action for the associated gravity theory and to derive the corresponding equations of motion.


\subsection{The algebraic setup}

In section~\ref{sec_gendiffg} we have reviewed how a Lie-algebroid structure
can give rise to a diffeomorphism invariant differential geometry framework. 
Unfortunately, the Koszul bracket \eqref{koszul}, which would be the first choice, does not provide a proper Lie algebroid  on $T^*M$ in the case of non-vanishing $R$-flux.
However, let us note the following: when translating the gauge symmetries from the tangent to the co-tangent bundle, in equation \eqref{qps} we have chosen the quasi-Poisson structure $\beta$ to be the inverse of the Kalb-Ramond field $B$. This allows us here to relate the $R$-flux to the $H$-flux in the following way
\eq{
	H_{abc} &= 3\,\partial_{[a}B_{bc]} \\
	&= - 3\, B_{[b|m}\,(\partial_{|a|}\beta^{mn}) \, B_{n|c]} \\
	&= 3\, B_{[a|k|}\,B_{b|m|}\,B_{c]n}\,D^k\beta^{mn}\\
	&= B_{ak}\,B_{bm}\,B_{cn}\, \Theta^{mnk} \;,
}
or, employing \eqref{redef}, we equivalently obtain
\eq{\label{RaH}
	\Theta^{abc} = \beta^{am}\,\beta^{bn}\,\beta^{ck}\,H_{mnk} \, .
}
A proper Lie algebroid structure on $T^*M$ can  be constructed using the
\emph{$H$-twisted Koszul bracket} which has appeared in this context for instance in \cite{Blumenhagen:2012pc}\hspace{2pt}\footnote{Note that the bracket  in \cite{Blumenhagen:2012pc} is defined with the opposite sign for the $H$-flux term. However, this difference can be removed by replacing $B\to-B$, which does not  change the properties of the bracket.}
\eq{\label{koszulH}
	[\xi,\eta]_K^H = [\xi,\eta]_K - \iota_{\beta^\sharp\eta}\iota_{\beta^\sharp\xi}H \;,
}
where $[\cdot , \cdot ]_K$ denotes the usual Koszul bracket \eqref{koszul}.
In this way, we obtain a  Lie-algebroid structure on $T^*M$ for an $R$-flux $\Theta^{abc}$ of the form \eqref{RaH}. Indeed, the corresponding Jacobi identity can be evaluated to 
\eq{\label{jacHK}
	\mathrm{Jac}_K^H(\eta,\chi,\zeta)
	= d\bigl(\mathcal{R}(\eta,\chi,\zeta)\bigr) + \iota_{(\iota_\zeta\iota_\chi\mathcal{R})}d\eta
		+ \iota_{(\iota_\eta\iota_\zeta\mathcal{R})}d\chi
		+ \iota_{(\iota_\chi\iota_\eta\mathcal{R})}d\zeta \;,
}
with
\eq{
	\mathcal{R}^{abc} = \Theta^{abc} - \beta^{am}\,\beta^{bn}\,\beta^{ck}\,H_{mnk} \, .
}
Thus, the Jacobiator \eqref{jacHK} vanishes upon setting $\mathcal{R}=0$, which implies \eqref{RaH}, and so we arrive at a proper Lie algebroid.
For later reference, let us also evaluate the $H$-twisted Koszul bracket on a basis $\{e^a\}$ of $\Gamma(T^*M)$ to obtain
\eq{\label{calQ}
	[e^a,e^b]_K^H &= \bigl(Q_c{}^{ab}-\beta^{am}\,\beta^{bn}\,H_{mnc} \bigr)\hspace{1pt}e^c \\
	&= \bigl(Q_c{}^{ab}+\Theta^{abm}\,\beta_{mc} \bigr)\hspace{1pt}e^c \\
	&= \mathcal{Q}_c{}^{ab}\hspace{1pt} e^c \;.
}
The Jacobi identity for this basis, which is the fifth Bianchi identity in \cite{Blumenhagen:2012pc} for $\mathcal{R}=0$, reads
\eq{\label{calQjac}
	0 = D^{[a}\mathcal{Q}_m{}^{bc]} + \mathcal{Q}_m{}^{[a|p}\,\mathcal{Q}_p{}^{|bc]} \;.
}

Furthermore, from equation \eqref{lieilldef} we can infer that the standard Koszul bracket of two $\beta$-tensors does not result in a $\beta$-tensor but includes an additional term proportional to the $R$-flux
\eq{
	\fa\delta_{\xi} \bigl([\eta,\chi]_K\bigr)_a
	= \bigl( \fa{\mathcal L}_{\xi}[\eta,\chi]_K \bigr)_a + \eta_m\,\chi_n\,\Theta^{mnk}\bigl(\hspace{1pt}d\xi\hspace{1pt}\bigr)_{ak}\,.
}
However, for the $H$-twisted bracket \eqref{koszulH} we compute
\eq{\label{HKbt}
	\fa\delta_{\xi} \bigl([\eta,\chi]_K^H\bigr)_a
	&= \fa\delta_{\xi}\bigl([\eta,\chi]_K\bigr)_a
		- \fa\delta_{\xi}\bigl(H_{abc}\,\beta^{bm}\hspace{1pt}\beta^{cn}\hspace{1pt}\eta_m\hspace{1pt}\chi_n\bigr) \\
	&= \fa\delta_{\xi} \bigl([\eta,\chi]_K\bigr)_a
		+ \fa\delta_{\xi}\bigl(\Theta^{mnk}\beta_{ka}\,\eta_m\,\chi_n\bigr) \\
	&= \bigl( \fa{\mathcal L}_{\xi}[\eta,\chi]_K^H \bigr)_a + \eta_m\hspace{1pt}\chi_n\hspace{1pt}
	\Theta^{mnk}(d\xi)_{ak}+\Theta^{mnk}\bigl(\fa\Delta_{\xi}\beta_{ka}\bigr)\eta_m\,\chi_n \\
	&= \bigl( \fa{\mathcal L}_{\xi}[\eta,\chi]_K^H \bigr)_a \,,
}
where we have used \eqref{RaH}. The $\beta$-variation of $\beta$ and the $R$-flux was given in   \eqref{tbeta} and \eqref{tR}, respectively, and we denoted $\fa\Delta_{\xi}=\fa\delta_\xi - \fa{\mathcal L}_\xi$.  Thus, contrary to the untwisted case, the $H$-twisted Koszul bracket of two $\beta$-tensors is again a $\beta$-tensor. 
Therefore, the Lie algebroid $(T^*M,[\cdot,\cdot]_K^H,\beta^\sharp;\mathcal{R}=0)$ provides a proper framework for describing a non-vanishing $R$-flux of the form \eqref{RaH} for a diffeomorphism invariant theory.


\subsection{Connection, torsion and curvature}

In this section,  we introduce a connection on $T^*M$ which covariantizes the
derivative $D$, and discuss in detail torsion and curvature tensors defined
with respect to the $H$-twisted Koszul bracket. Note that invariance under
diffeomorphisms  is intrinsic to our constructions, as  \eqref{koszulH} gives a proper Lie algebroid for an $R$-flux $\Theta$ of the form \eqref{RaH}. Furthermore, we will see that also $\beta$-tensoriality is manifest which is mainly due to \eqref{HKbt}.
In the following, we therefore assume that all tensors are $\beta$-tensors if not otherwise specified.


\subsubsection*{Connection}

As we have discussed in section~\ref{sec_gendiffg}, a connection on $T^*M$ which covariantizes the differential \eqref{D} is given by a $\mathcal C^{\infty}$-linear map $\fa\nabla:\Gamma(T^*M)\times \Gamma(T^*M)\to\Gamma(T^*M)$ satisfying the Leibniz rule \eqref{covder}. In the present context, this implies that 
\eq{\label{connL}
	\fa\nabla_{\!\xi}(f\hspace{1pt}\eta) &= \bigl( (\beta^\sharp\xi)f\bigl) \hspace{1pt}\eta
		+ f\,\fa\nabla_{\!\xi} \eta \\
	&= \xi_m (D^m f ) \hspace{1pt}\eta + f\,\fa\nabla_{\!\xi}\eta \;,
}
with $f$ a function and $\xi$, $\eta$ one-forms.
In local coordinates, the covariant derivative can be characterized as follows. For a frame $\{e^a\}$ of $\Gamma(T^*M)$ we introduce connection coefficients $\fa\Gamma_c{}^{ab}$ by
\eq{
	\fa\nabla_{\!e^a}e^b \equiv \fa\nabla^a\,e^b = \fa\Gamma_c{}^{ab}\, e^c \;,
}
and  using \eqref{connL}  we obtain
\eq{\label{cd1}
	\fa\nabla^a\eta_b = D^a\eta_b + \fa\Gamma_b{}^{am}\,\eta_m \,.
}
Requiring compatibility of the connection with the insertion, that is $	D^a(\iota_X\eta) = \iota_X(\fa\nabla^a\eta)+\iota_\eta(\fa\nabla^a X)$,
for a vector field $X$ we then find
\eq{
	\fa\nabla^a X^b = D^a X^b - \fa\Gamma_m{}^{ab}\,X^m \,.
}
Generalizing these expressions, we obtain  the following rule for applying the covariant derivative to an $(r,s)$-tensor 
\eq{
	\fa\nabla^c\,T_{a_1\dots a_r}{}^{b_1\dots b_s}
	= D^c\, T_{a_1\dots a_r}{}^{b_1\dots b_s}
	&+\sum_{i=1}^r\fa\Gamma_{a_i}{}^{cm}\,
		T_{a_1\dots a_{i-1}ma_{i+1}\dots a_r}{}^{b_1\dots b_s}\\
	&-\sum_{i=1}^s\fa\Gamma_{m}{}^{cb_i}\,
		T_{a_1\dots a_r}{}^{b_1\dots b_{i-1}mb_{i+1}\dots b_s} \,.
}

As we have discussed in section~\ref{sec:betaten}, if a function $f$ is a $\beta$-scalar then $D^a f$ is a $\beta$-tensor, which by definition also includes  tensoriality under usual diffeomorphisms. Now, 
the anomalous diffeomorphism and $\beta$-diffeomorphism transformations  $\Delta_X=\delta_X-L_X$ and $\fa\Delta_{\xi}=\fa\delta_{\xi}-\fa{\mathcal L}_{\xi}$ of  $D^a\eta_b$ 
can be computed as
\eq{\label{andeta}
	\Delta_X(D^a \eta_b) = D^a(\partial_bX^m) \eta_m\,, \hspace{25pt}
	\fa\Delta_{\xi}\,(D^a \eta_b) = -D^a(D^m\xi_b
		- \xi_n\,\mathcal{Q}_b{}^{nm}) \hspace{1pt}\eta_m\, .
}
Thus, $D^a\eta_b$ does not behave as a $\beta$-tensor 
and so the connection coefficients $\fa\Gamma_c{}^{ab}$ have to transform anomalously 
to compensate for \eqref{andeta}. In particular, we have to require
\eq{\label{antcc}
	\Delta_X\fa\Gamma_c{}^{ab} = -D^a(\partial_cX^b) \;, \hspace{40pt}
	\fa\Delta_{\xi}\,\fa\Gamma_c{}^{ab} = D^a(D^b\xi_c
		- \xi_m\,\mathcal{Q}_c{}^{mb}) \; ,
}
in order for the covariant derivative \eqref{cd1} to behave as a $\beta$-tensor. A similar observation can be made for $D^a X^b$, so that with \eqref{antcc}  the covariant derivative correctly maps $\beta$-tensors to $\beta$-tensors. However, let us note that for the Levi-Civita connection to be introduced below, it can be checked explicitly that the connection coefficients indeed transform as \eqref{antcc}.


\subsubsection*{Torsion}

The general expression for the torsion  $T\in\Gamma(\wedge^2TM\otimes T^*M)$ has been given in \eqref{curv}, and for the $H$-twisted Koszul bracket it reads
\eq{\label{torsion}
  \fa T(\xi,\eta) = \fa\nabla_{\!\xi}\,\eta-\fa\nabla_{\!\eta}\, \xi-[\xi,\eta]_K^H \;,
}
which is $C^\infty(M)$-linear in both arguments since $\beta^\sharp$ is an algebra-ho\-mo\-mor\-phism. Note that this property would fail for the un-twisted Koszul bracket. 
Furthermore, with the connection well-defined on $\beta$-tensors and with \eqref{HKbt}, the torsion \eqref{torsion}  also is a $\beta$-tensor. 
Locally,  \eqref{torsion} can be written as
\eq{
	\fa T_c{}^{ab} = \iota_{e_c}\fa T(e^a,e^b)
	= \fa\Gamma_c{}^{ab} -\fa\Gamma_c{}^{ba} - \mathcal{Q}_c{}^{ab} \,,
}
where $\mathcal Q_{c}{}^{ab}$ had been computed in \eqref{calQ}.
The anomalous transformation behavior of $\mathcal{Q}$ can be obtained directly from \eqref{tbeta} and \eqref{calQ} giving
\eq{
	\Delta_X\mathcal{Q}_c{}^{ab} &= -2\,D^{[a}(\partial_cX^{b]}) \;, \\
	\fa\Delta_{\xi}\,\mathcal{Q}_c{}^{ab} &= \mathcal{Q}_m{}^{ab}\,D^m\xi_c
	 + 2\,\mathcal{Q}_c{}^{m[a}\,D^{b]}\xi_m - 2\,\xi_m\,D^{[a|}\mathcal{Q}_c{}^{m|b]} \;,
}
which cancels the anomalous transformation of the anti-symmetrization of $\fa\Gamma_c{}^{ab}$. Thus also in components we see that the torsion \eqref{torsion} is a $\beta$-tensor if the connection coefficients transform as \eqref{antcc}.


\subsubsection*{Levi-Civita connection}

As already discussed in section~\ref{sec_gendiffg}, similar to standard differential geometry we can determine a unique connection by requiring metric compatibility and vanishing torsion. More concretely, for a metric $\fa g\in\Gamma(TM\otimes_{\mathrm{sym}}TM)$ let us require
\eq{
	(\beta^\sharp\xi)\hspace{1pt}\fa g(\eta,\chi) = \fa g\bigl(\fa\nabla_{\!\xi} \eta,\chi\bigr)
		+ \fa g\bigl(\eta, \fa\nabla_{\!\xi}\chi\bigr) \;,
}
and from \eqref{torsion} we see that vanishing torsion implies
\eq{
	\fa\nabla_{\!\xi}\eta-\fa\nabla_{\!\eta}\xi = [\xi,\eta]_K^H \;.
}
Employing these relations, we arrive at the  Koszul formula \eqref{Koszulformula}
\eq{\label{kkoszul}
	&\fa g\bigl(\fa\nabla_{\!\xi}\eta,\chi\bigr)
	= \frac{1}{2}\Bigl((\beta^\sharp\xi)\fa g(\eta,\chi)
		+(\beta^\sharp\eta)\fa g(\chi,\xi)-(\beta^\sharp\chi)\fa g(\xi,\eta) \\
	&\hspace{100pt} + \fa g\bigl([\xi,\eta]_K^H,\chi\bigr)+\fa g\bigl([\chi,\xi]_K^H,\eta\bigr)
		- \fa g\bigl([\eta,\chi]_K^H,\xi\bigr)\Bigr) \;,
}
which uniquely determines the Levi-Civita connection in the present context. By a slight abuse of notation, the latter will be denoted by $\fa\nabla$  from now on. Inserting then basis sections $\{e_a\}$ into \eqref{kkoszul},  the connection coefficients are determined as
\eq{\label{klc}
	\fa\Gamma_c{}^{ab} = \frac{1}{2}\,\fa g_{cm}\bigl(D^a\fa g^{bm}
	+D^b\fa g^{am}-D^m\fa g^{ab}\bigr)
	-\fa g_{cm}\,\fa g^{(a|n}\,\mathcal{Q}_n{}^{|b)m} + \frac{1}{2}\,\mathcal{Q}_c{}^{ab}\,.
}
Note that this Levi-Civita connection is not symmetric in the upper indices but has an anti-symmetric contribution from the last term in \eqref{klc}. Furthermore, from \eqref{kkoszul}  it is clear that 
\eqref{klc} has the expected transformation behavior \eqref{antcc}, which can also be checked explicitly.


\subsubsection*{Curvature}

On general grounds, in section~\ref{sec_gendiffg} the curvature $\fa R\in\Gamma(\wedge^2TM\otimes\mathrm{End}(T^*M))$ has been de\-fined by equation \eqref{curv}. For the present situation of the $H$-twisted Koszul bracket, this implies
\eq{
	\fa R(\eta,\chi)\xi = \bigl[\fa\nabla_{\!\eta},\fa\nabla_{\!\chi}\bigr]\xi
		- \fa\nabla_{\![\eta,\chi]_K^H}\, \xi \;,
}
which in components reads
\eq{
  \label{kriemt}
	\fa R_a{}^{bcd}\equiv \iota_{e_a}\bigl(\fa R(e^c,e^d)e^b\bigr)
	= 2\bigl(D^{[c}\fa\Gamma_a{}^{d]b} + \fa\Gamma_a{}^{[c|m}\,\fa\Gamma_m{}^{|d]b}\bigr)
		- \fa\Gamma_a{}^{mb}\,\mathcal{Q}_m{}^{cd} \, .
}
Since  the covariant derivative and the bracket give $\beta$-tensors,
also $\fa R$ is a $\beta$-tensor.
Using then the Bianchi identity \eqref{calQjac}
and raising indices with the metric $\fa g^{ab}$, we can show that the curvature with respect to the Levi-Civita connection \eqref{klc} admits the same symmetries and Bianchi identities as the usual curvature tensor, that is
\eq{\label{Rsymm}
	\fa R^{abcd} = -\fa R^{bacd} \;,\hspace{40pt}
	\fa R^{abcd} = -\fa R^{abdc} \;, \hspace{40pt}
	\fa R^{abcd} = \fa R^{cdab} \;,
}
as well as
\eq{
\label{Rsymm2}
       \fa R^{abcd}+\fa R^{adbc}+\fa R^{acdb}&=0 \;,\\
       \fa\nabla^m \fa R^{abcd}+\fa\nabla^d \fa R^{abmc}+\fa\nabla^c \fa
  		R^{abdm}&=0\;.
}
The Ricci tensor is defined  by $\fa R^{ab} = \fa R_m{}^{amb}$, which is symmetric in its indices due to \eqref{Rsymm}. In terms of the connection, it can be written as
\eq{
	\fa R^{ab} = D^m\fa\Gamma_m{}^{ba} - D^b\fa\Gamma_m{}^{ma}
		+ \fa\Gamma_n{}^{ba}\,\fa\Gamma_m{}^{mn}
		- \fa\Gamma_n{}^{ma}\,\fa\Gamma_m{}^{nb} \, .
}
Finally the Ricci scalar $\fa R = \fa g_{ab}\fa R^{ab}$ can be expanded in terms of the metric and the derivative $D^a$ in the following way
\eq{
\label{Rexpand}
     \hat R=-\Bigl[ &\hspace{15pt}D^a D^b \hat g_{ab}
          - D^a\left( \hat g_{ab} \, \hat g^{mn}\,  D^b \hat g_{mn}  \right)\\
          &-{1\over 4} \hat g_{ab}\Bigl( D^a  \hat g_{mn}\, D^b  \hat g^{mn}
                            -2 D^a  \hat g_{mn}\, D^m  \hat g^{nb}
           -  \hat g_{mn}\,  \hat g_{pq}\, D^a  \hat g^{mn}\, D^b  \hat g^{pq}\Bigr)\\
      &+{1\over 4}  \hat g_{ab}\,  \hat g_{mn} \, \hat g^{pq}\,
		{\cal Q}_p{}^{ma} {\cal Q}_q{}^{nb}
      +{1\over 2}  \hat g_{ab}\, {\cal Q}_m{}^{nb}\, {\cal Q}_n{}^{ma}
     +  \hat g_{ab}\, {\cal Q}_m{}^{ma}\, {\cal Q}_n{}^{nb} \\
      &+2   D^a \bigl( \hat g_{ab} \, {\cal Q}_m{}^{mb} \bigr)
      -  \hat g_{ab}\, \hat g_{mn}\,  D^a \hat g^{pn}  \,   {\cal Q}_p{}^{bm}
     +   \hat g_{ab} \,\hat g^{mn}\,  D^a \hat g_{mn}   \,  {\cal
       Q}_p{}^{bp}\,\Bigr]\;.
}


\subsubsection*{Summary}

In this section we have seen that the Lie algebroid on the co-tangent bundle defined by the $H$-twisted Koszul bracket can be used to formulate  the usual geometric objects in a manifest $\beta$-tensorial way. This excels this framework as the one suitable for incorporating both transformations into a geometric setup. With  the relevant notions at hand, we are now able to formulate a gravity theory on $T^*M$.


\subsection{Symplectic gravity}

In this section, we  construct an Einstein-Hilbert action invariant under standard as well as $\beta$-diffeomorphisms, which we call \emph{bi-invariant} for short.
This action contains the metric $\hat{g}^{ab}$, a bi-vector ${\beta}^{ab}$ and a dilaton $\phi$ as dynamical fields.


\subsubsection*{Invariant action}

As we have illustrated in the last section, it is possible to construct a Ricci scalar $\fa{R}$ which behaves as a scalar with respect to both types of diffeomorphisms.
Furthermore, by construction, the derivative of the dilaton $D^a \phi$ is a $\beta$-tensor (thus a standard tensor in particular) and therefore the corresponding kinetic term $\hat{g}_{ab} D^a \phi D^b\phi$ behaves as a $\beta$-scalar. Also, the $R$-flux $\Theta^{abc}$ is tensorial with respect to $\beta$-diffeomorphisms as was shown in \eqref{tR}, and it behaves as a standard tensor due to its definition \eqref{flux_78} in terms of the Schouten-Nijenhuis bracket of ${\beta}$ with itself. 
Therefore, the following Lagrangian is a scalar with respect to both types of diffeomorphisms
\eq{
   \hat{\cal L}=e^{-2\phi} \left(   \hat R -{1\over 12} \Theta^{abc}\, \Theta_{abc}
              +4\hspace{0.5pt} \hat g_{ab}\, D^a\phi D^b \phi\right)\;.
}
This Lagrangian has been constructed in such a way to resemble the bosonic low-energy effective action \eqref{stringaction_intro}. Analogous to the geometric case, $\Theta$ can also be included as (con-)torsion of the connection.

To obtain a bi-invariant action, we have to find an appropriate measure $\mu$. More precisely, the variation of 
\eq{
     \hat S={1\over 2\kappa^2} \int d^nx\, \mu(\fa g,  \beta)\, \hat{\cal L}\, ,
}
under standard and $\beta$-diffeomorphisms has to give a total derivative. As it turns out, the direct analogue to Riemannian geometry, namely the measure $\mu = \sqrt{-|\hat{g}|}$ with $|\fa g|=\det \fa g^{ab}$, does not lead to the desired result. This can be seen from
\eq{\label{varmu}
\delta_X  \bigl(\sqrt{-|\fa g|}\,\fa{\cal L}\hspace{1pt}\bigr) &= \partial_m\bigl(X^m \sqrt{-|\fa g|}\,\fa {\cal L} \hspace{1pt}\bigr) -2\sqrt{-|\fa g|}\, \fa{\cal L}\, (\partial_m X^m) \, ,\\[3pt]
\fa\delta_{\xi}\bigl(\sqrt{-|\fa g|}\,\fa{\cal L}\hspace{1pt}\bigr)&=
       \partial_m\bigl( \sqrt{-|\fa g|}\, \fa {\cal L}\, 
       \xi_n\,\bigr)\beta^{nm}
      -  \sqrt{-|\fa g|}\, \fa {\cal L}\,  \xi_m (\partial_n
       \beta^{mn}) \, .
   }
Obviously, the right-hand sides in \eqref{varmu} are not total derivatives which would be required for the action to be invariant. However, taking as an additional factor the determinant of  $ \beta^{-1}$ into account, that means
\eq{\label{correctmeasure}
\mu =  \sqrt{-|\fa g|}\, \bigl| \beta^{-1}\bigr| \;,
}
we obtain the correct behavior under both types of diffeomorphisms. This can be seen by considering the variation of the determinant of the bi-vector
\eq{\label{varbeta}
\delta_X  \bigl|  \beta^{-1} \bigr| &= X^m \partial_m \bigl| \beta^{-1}\bigr| + 2\hspace{1pt}\bigl| \beta^{-1} \bigr| \partial_m X^m \;, \\
\fa \delta_{ \xi} \bigl| \beta^{-1}\bigr| &= 2 \bigl| \beta^{-1}\bigr|\, \xi_m\, \partial_n \beta^{mn} + \xi_m \beta^{mk} \partial_k \bigl| \beta^{-1}\bigr| \;,
}
so that  the combination of \eqref{varmu} and \eqref{varbeta} results in a total derivative. We therefore propose the following bi-invariant Einstein-Hilbert action coupled to a dilaton $\phi$ and $R$-flux $\Theta^{abc}$
\eq{
\label{finalaction_pre}
      \hat S={1\over 2\kappa^2} \int d^nx\, \sqrt{-|\hat g|}\,\bigl|  \beta^{-1} \bigr|\, e^{-2\phi}
     \Bigl(   \hat R -{1\over 12} \Theta^{abc}\, \Theta_{abc}
              +4\hspace{1pt} \hat g_{ab}\, D^a\phi D^b \phi\Bigr)\, .
}
Due to the appearance of the \mbox{(quasi-)}symplectic structure $\beta^{ab}$, we will call the theory defined by the action \eqref{finalaction_pre} \emph{symplectic gravity}.


\subsubsection*{Remarks}

Let us close this section with two remarks about the  measure \eqref{correctmeasure}.
\begin{itemize}

\item In general, the determinant of an anti-symmetric matrix vanishes in odd
  dimensions. Thus, our measure \eqref{correctmeasure} only makes sense for
  even dimensions, e.g. for symplectic manifolds. For the  latter case
  one has 
\eq{
  \det \beta_{ab} = \bigl( {\rm Pfaff}\, \beta_{ab} \bigr)^2 \;,
}
so that the determinant $| \beta^{-1} |$ is always non-negative.

\item In the Lie-algebroid construction of section~\ref{sec_betadiffeos}, we have effectively replaced the tangent bundle of a manifold by the co-tangent bundle. Performing the same procedure for 
an integral, we would formally obtain
\eq{ \label{newint}
\int \sqrt{-|G|}\,dx^1 \wedge \ldots \wedge dx^{n} \quad\to\quad\int
\sqrt{-|\fa g|}\,\partial_1 \wedge \ldots \wedge \partial_{n}\, .
}
Employing then the inverse of the anchor, we can relate the right-hand side to a standard integral by using $\partial_a = \beta_{ab}\, dx^b$ which results in the same measure as in \eqref{correctmeasure}
\eq{
\int \sqrt{-|\fa g|}\, \partial_1 \wedge \dots \wedge \partial_{n} = \int \sqrt{-|\fa g|} \hspace{1pt}\bigl|  \beta^{-1} \bigr|\, dx^1 \wedge \ldots \wedge dx^n \,.
}
However, let us
note again that this replacement is only possible in an even number of dimensions, otherwise the determinant of $\beta$ would vanish and the anchor would not be invertible.

\end{itemize}


\subsection{Equations of motion}
\label{sec_the_real_eom}

After having derived the action \eqref{finalaction_pre} for the symplectic gravity theory, we now turn to the resulting equations of motion for the metric $\fa g^{ab}$, the bi-vector $ \beta^{ab}$ and the dilaton $\phi$.
Although being straightforward, the computation turns out to be rather involved. We therefore only provide some details on the major steps of the calculation as well as  some important formulas.

First, we note that by using the explicit form \eqref{klc} of the connection coefficients $\fa\Gamma_c{}^{ab}$, one can check  the following relation for an arbitrary one-form $\eta_a$
\eq{
  \int d^nx\, \sqrt{-|\hat g|}\hspace{2pt}  \bigl| \beta^{-1} \bigr|\, \fa \nabla^a \eta_a 
  = -\int d^nx\,\, \partial_a\Bigl(\sqrt{-|\hat g|}\hspace{2pt}  \bigl| \beta^{-1} \bigr|\, \fa \beta^{am}\, \eta_m\Bigr) = 0
}
for a manifold without boundary.
Employing this formula, the variation of  \eqref{finalaction_pre} with respect to the dilaton $\phi$ can  easily be performed. Setting  to zero the variation, we obtain the  equation of motion
\eq{
  \label{eom_01}
  \mbox{ I$\hspace{0.5pt}'$ :} \hspace{30pt}
  0 =  \hat R -{1\over 12}\hspace{1pt} \Theta^{abc}\, \Theta_{abc}
   -4\hspace{0.5pt} \hat g_{ab}\hspace{1pt} \fa\nabla^a\phi \fa\nabla^b \phi
   +4\hspace{0.5pt} \hat g_{ab}\hspace{1pt} \fa\nabla^a \fa\nabla^b \phi \;.
}
Next, for the variation of the action with  respect to the metric $\fa g^{ab}$, we note the   Palatini identity for the Ricci tensor $\fa R^{ab}$
\eq{
  \delta_{\fa\Gamma} \fa R^{ab} = \fa \nabla^m \delta \fa\Gamma_m{}^{ab}
   - \fa\nabla^a \delta \fa\Gamma_m{}^{mb}
   + \fa\Gamma_n{}^{mb} \bigl( \delta \fa\Gamma_m{}^{na} - \delta\fa\Gamma_m{}^{an} \bigr) \;.
}
Using then again the explicit form of the connection coefficients $\fa\Gamma_c{}^{ab}$ and setting to zero the variation of the action, we arrive at
\eq{
  \label{eom_02}
  \mbox{ II :} \hspace{30pt}
  0 = \fa R^{ab} + 2 \hspace{0.5pt} \fa\nabla^a \fa\nabla^b \phi - \frac14 \hspace{0.5pt} \Theta^{amn}
  \Theta^b{}_{mn} - \frac12 \, \fa g^{ab} \bigl[ \, \mbox{$\phi$ eom} \, \bigr] \;,
}
where the last term vanishes due to the equation of motion \eqref{eom_01} for $\phi$.
The variation of the action \eqref{finalaction_pre} with respect to the
bi-vector  is a more involved task, as  $\beta^{ab}$ appears for instance in all derivatives $D^a$. However, setting again to zero the variation, we obtain
\eq{
  \mbox{ III :} \hspace{30pt}
  0 = \frac12 \hspace{1pt} \fa\nabla^m \Theta_{mab} - (\fa\nabla^m \phi) \hspace{1pt}\Theta_{mab}
  &+ 2 \hspace{1pt} \fa g_{ap} \beta_{bq}  \bigl[ \, \mbox{$\fa g$ eom} \, \bigr] ^{pq} \\
  &+ \beta_{ab}  \bigl[ \, \mbox{$\phi$ eom} \, \bigr]  \;,
}
where the last two terms vanish because of \eqref{eom_02} and \eqref{eom_01}, respectively.
Finally, we note that the trace of the equation of motion for $\fa g$ reads as follows
\eq{
  \label{eom_03}
  \mbox{ II$\hspace{0.5pt}'$ :} \hspace{30pt}
  0 = \fa R + 2 \hspace{0.5pt} \fa g_{ab} \fa\nabla^a \fa\nabla^b \phi - \frac14 \hspace{0.5pt} \Theta^{abc}
  \Theta_{abc} \;.
}
Combining then \eqref{eom_03} with \eqref{eom_01}, we arrive at the following set of independent equations of motion for the metric, bi-vector and dilaton 
\eq{
  \label{eom_final}
  \mbox{ I :} \hspace{30pt}
  &0 =
   -\frac12\hspace{0.5pt} \hat g_{ab}\hspace{1pt} \fa\nabla^a \fa\nabla^b \phi
   +\hat g_{ab}\hspace{1pt} \fa\nabla^a\phi \fa\nabla^b \phi
    -\frac{1}{24}\hspace{1pt} \Theta^{abc}\, \Theta_{abc} \;, \\
  \mbox{ II :} \hspace{30pt}
  & 0 = \fa R^{ab} + 2 \hspace{0.5pt} \fa\nabla^a \fa\nabla^b \phi
  - \frac14 \hspace{0.5pt} \Theta^{amn}
  \Theta^b{}_{mn} \;, \\
  \mbox{ III :} \hspace{30pt}
  &0 = \frac12 \hspace{1pt} \fa\nabla^m \Theta_{mab} - (\fa\nabla^m \phi) \hspace{1pt}\Theta_{mab}
  \;.
}
Let us emphasize that these expressions take the same form as the well-known
formulas in the standard setting, if one performs the replacements $\fa g^{ab}
\to G_{ab}$, $\fa\nabla^a \to \nabla_a$, $\Theta^{abc} \to H_{abc}$ and $\fa
R^{ab} \to R_{ab}$. 
In section \ref{sec_eom}, some simple solutions to these
equations will be discussed.


\section{Relations to string theory}
\label{sec_relst}

In the previous sections we have developed a generalized differential-geometry
framework based on the theory of Lie algebroids,
which led to the bi-invariant action\hspace{2pt}\footnote{In order to clearly distinguish between objects in standard and symplectic frame, in the present and subsequent sections  we use a hat not only for the metric $\fa g^{ab}$ but also for the bi-vector $\fa\beta^{ab}$ and the $R$-flux $\fa\Theta^{abc}$.}
\eq{
\label{finalaction}
      \hat S= \frac{1}{2\kappa^2} \int d^nx\, \sqrt{-|\hat g|}\: \bigl|\hat \beta^{-1}\bigr|\: e^{-2\phi}
      \Bigl(   \hat R -\tfrac{1}{12} \hspace{1pt}\fa\Theta^{abc}\, \fa\Theta_{abc}
              +4\hspace{1pt} \hat g_{ab}\, D^a\phi D^b \phi\Bigr)\, ,
}
for the metric $\fa g^{ab}$, a (quasi-)symplectic two-vector $\fa \beta^{ab}$ and the dilaton $\phi$.
In this section, we  clarify the relation between \eqref{finalaction} and
the low-energy effective action for the massless modes of the bosonic string
\eq{
\label{stringaction}
S={1\over 2\kappa^2}
\int
\hspace{-0.75pt}
d^nx \hspace{1pt}\sqrt{-|G|}\hspace{1pt} e^{-2\phi}\Bigl(R-{\textstyle{1\over
    12}} H_{abc} H^{abc}
+4 \hspace{0.5pt} G^{ab}\, \partial_a \phi \hspace{1pt}\partial_b \phi
 \Bigr) \;,
}
where the latter of course also describes the massless modes in the
NS-NS sector of  type II superstring theories.
We will see that the two actions \eqref{finalaction} and \eqref{stringaction}
are related by a change of fields  from the $(G,B,\phi)$-frame
to the $(\fa g, \fa \beta,\phi)$-frame. 
Furthermore, by extending the field redefinition from the
NS-NS sector to the RR (space-time bosons) and NS-R sectors (space-time fermions),
we will  propose the form of a symplectic type IIA supergravity theory.
Even though the construction of such a supersymmetric action from first
symmetry principles is beyond the scope of this paper, we
expect it to involve a super-symmetrization of the $\beta$-diffeomorphisms.


\subsection{Effective action for the bosonic string}

From results in generalized geometry and double field
theory, one would expect that
the relation between the geometric and non-geometric fields is given by
\eq{
    \label{relation_dft}
      \tilde g &= \hphantom{-}(G+ B)^{-1}\, G\,  (G-B)^{-1} \;, \\
\tilde\beta &=- (G+B)^{-1}\, B\,  (G- B)^{-1}\;.
}
However, as the computation in \cite{Andriot:2012wx,Andriot:2012an} shows,
starting from the action \eqref{stringaction} and
inserting (the inverse of) this transformation
does not lead to \eqref{finalaction}.
But, a second natural possibility for a change of fields
arises by observing that the relation between
$(G,B)$ and $(\tilde g,\tilde\beta)$ is formally the same
as in the study of D-branes
in two-form flux backgrounds. In particular, in the Seiberg-Witten limit \cite{Seiberg:1999vs}, that is where
a brane theory with flux is effectively described
by a non-commutative gauge theory, the relation between
the fields is given by the formulas\hspace{2pt}\footnote{Note that we are not taking a true limit  $G\to 0$, implying
that we are not neglecting  any terms from the action.}
\eq{
\label{redefine}
    B=\hat\beta^{-1}  \, ,\hspace{40pt}
    G = -\hat\beta^{-1}\, \hat g\, \hat\beta^{-1} \; ,
}
which in components reads 
\eq{ 
   B_{ab}=  \fa\beta_{ab}\, ,
   \hspace{40pt} G_{ab}= \fa\beta_{am}\,
\fa\beta_{bn}\, \fa g^{mn} \, ,
}
with  $\fa\beta_{ab}=( \fa\beta^{-1} )_{ab}$.
This change of fields of course is the same as the one we have introduced in equation \eqref{redef}.


\subsubsection*{Relation between actions}

We now show that the two actions \eqref{stringaction} and \eqref{finalaction} are related by the field redefinition \eqref{redefine}. We therefore first consider
the determinant of the metric $G$ for which we compute
\eq{
      \sqrt{-|G|}=\sqrt{-|\fa g|}\: \bigl|\fa\beta^{-1}\bigr| \;.
}
Next, the transformation of the Christoffel connection $\Gamma^c{}_{ab}$ in the standard frame under the change of fields \eqref{redefine} is found to be of the following form
\eq{
\label{transgamma}
\Gamma^c{}_{ab}=-\fa \beta^{cp}\, \fa \beta_{am}\, \fa \beta_{bn}\,
         \fa \Gamma_p{}^{mn} - \fa\beta_{nb} \, \partial_a \fa\beta^{cn}\, ,
}
where $\fa\Gamma_p{}^{mn}$ was given in \eqref{klc}.
Employing this result, we can determine the behavior of the usual Riemann curvature tensor $R^d{}_{cab}$
under the above transformations as
\eq{
\label{curvaturetrafo}
     R^d{}_{cab}=-\fa\beta^{dq}\, \fa\beta_{cp}\,
    \fa\beta_{am}\, \fa\beta_{bn}\,
      \fa R_q{}^{pmn}     \, ,
}
with $\fa R_q{}^{pmn}$ as defined in  \eqref{kriemt}.
The Ricci tensor tensor and Ricci scalar are then computed in the following way
\eq{
  R_{ab}=  \fa\beta_{am} \, \fa\beta_{bn} \, \fa R^{mn}\, , \hspace{40pt}
  R=\fa R\, .
}
Next, we turn to the field strength of the Kalb-Ramond field. Recalling the  convention $H_{abc} = 3\hspace{1pt} \partial_{[a} B_{bc]}$, under the transformation \eqref{redefine} it behaves as
\eq{
    H_{abc}=\fa\beta_{am}\, \fa\beta_{bn}\, \fa\beta_{cp}\,  \fa\Theta^{mnp} \;,
}
which implies that $H_{abc} H^{abc}=\fa\Theta^{abc}\hspace{0.5pt}\fa\Theta_{abc}$.
And  since the dilaton $\phi$ is invariant under the field redefinition, we can write
\eq{
      \partial_a\phi = \fa\beta_{am} D^m\phi \;,
}
and so the corresponding kinetic term transforms as expected.
Therefore, collecting these results, we can show that indeed the action  \eqref{stringaction}
is related to \eqref{finalaction} via the field redefinition \eqref{redefine}, that is
\eq{
S\bigl(\,G(\fa g,\fa\beta),\,B(\fa g,\fa\beta),\,\phi\,\bigr)=\hat S\bigl(\fa g,\fa\beta,\phi\bigr)\, .
}


\subsubsection*{Higher-order corrections}

The effective action \eqref{stringaction} for the massless string modes is known to receive
higher-order $\alpha'$-corrections. Due to the freedom of
field redefinitions these are not unique, however,
all the terms appearing at next to leading order
 \cite{Metsaev:1987bc,Metsaev:1987zx,Hull:1987yi} 
can be expressed in terms of  covariant derivatives of
the curvature tensor $R_{abcd}$, the
three-form $H_{abc}$ and the dilaton $\partial_a \phi$.
Since we have determined how each of these building blocks transforms
under \eqref{redefine}, we have a well-motivated  guess for the
form of the higher-order corrections in the symplectic gravity frame.
For instance,   the next to leading order corrections 
to the bosonic string effective action are expected to take the form
\eq{
 &  \fa S^{(1)} =  \frac{1}{2\kappa^2}
    \,\frac{\alpha'}{4}
  \int d^{26}x\, \sqrt{-|\fa g|}\, \bigl|\fa\beta^{-1}\bigr|\, e^{-2\phi}
   \Bigl( \fa R^{abcd}\, \fa R_{abcd}
     -{\textstyle {1\over 2}}  \fa R^{abcd}\, \fa \Theta_{abm}
     \fa\Theta_{cd}{}^m\\
   &\hspace{142pt}
      + {\textstyle{1\over 24}}   \fa \Theta_{abc}\, \fa \Theta^{a}{}_{mn}\, \fa \Theta^{bm}{}_p\,
        \fa \Theta^{cnp}
    -{\textstyle {1\over 8}} (\fa\Theta^2)_{ab} \, (\fa\Theta^2)^{ab}     \Bigr) \,,
}
where we have abbreviated $(\fa\Theta^2)_{ab}=\fa\Theta_{amn}\, \fa\Theta_b{}^{mn}$.


\subsubsection*{Remarks}

Let us conclude this section with some remarks.
\begin{itemize}

\item The above results suggest that objects which are tensors with respects to both
types of diffeomorphisms show a simple transformation behavior under
the field redefinition \eqref{redefine}. In particular, indices are ``raised and lowered''
by contracting with the bi-vector $\fa\beta$
\eq{
\label{nicetrafon}
     T_{a_1\ldots a_n}=\fa\beta_{a_1 b_1}\ldots \fa\beta_{a_n b_n}\,
     \fa T^{b_1\ldots b_n}  \;.
}
More concretely, by explicit computation one shows that if \raisebox{0pt}[\ht\strutbox][4pt]{$\delta^{\rm gauge}_{\fa\xi} T_{a_1\ldots a_n}=0$}, where 
\raisebox{0pt}[\ht\strutbox][4pt]{$\delta^{\rm gauge}_{\fa \xi}$}
has been defined in \eqref{gtbg}, then \raisebox{0pt}[\ht\strutbox]{$\fa T^{b_1\ldots b_n}$} transforms as a $\beta$-tensor. That means, if $T_{a_1\ldots a_n}$ is invariant under $B$-field gauge transformations in the standard $(G,B)$-frame, then $\fa T^{b_1\ldots b_n}$ behaves as a $\beta$-tensor in the symplectic $(\fa g, \fa\beta)$-frame.

\item Furthermore, one can show that if a tensor $T_{a_1\ldots a_n}$ transforms
as  in equation \eqref{nicetrafon}, then its covariant derivative also satisfies
\eq{
\label{covderivtrafo}
        \nabla_a T_{a_1\ldots a_n} = \fa\beta_{a b}\,
      \fa\beta_{a_1 b_1}\ldots \fa\beta_{a_n b_n} \fa\nabla^b \fa T^{b_1\ldots
        b_n}\,,
}
where both connections are the Levi-Civita connections in the corresponding frame.

\item When contracting formula \eqref{curvaturetrafo} with the appropriate metric, it can be brought into the following form
\eq{
\label{curvaturetrafotwo}
    R_{dcab}=\fa\beta_{dq}\, \fa\beta_{cp}\,
    \fa\beta_{am}\, \fa\beta_{bn}\,
      \fa R^{qpmn}  \, .
}
Using then \eqref{covderivtrafo}, the symmetries and two Bianchi identities
of the Riemann tensor $\fa R^{pqmn}$ follow immediately from the properties of $R_{dcab}$
\eq{
  \label{bianchis}
   &  \fa R^{pqmn}= -\fa R^{pqnm} = -\fa R^{qpmn} =  \fa R^{mnpq}  \;, \\[0.1cm]
   &  0 = \fa R^{pqmn}+\fa R^{pmnq}+\fa R^{pnqm} \;,  \\
   &0 =   \fa\nabla^k \fa R^{pqmn}+\fa\nabla^m \fa R^{pqnk}+\fa\nabla^n \fa
  R^{pqkm} \;.
}
Recall that using a direct approach, these relations  have already been encountered in \eqref{Rsymm} and \eqref{Rsymm2}.

\item Above we have shown that the two actions \eqref{finalaction}
and \eqref{stringaction} are related via  the field redefinition \eqref{redefine}.
As a consequence, we can infer that the action which appeared
in \cite{Andriot:2012wx,Andriot:2012an}
is related to \eqref{finalaction}
via
\eq{
  \hat\beta =\tilde\beta - \tilde g \tilde\beta^{-1} \tilde g \;,
  \hspace{40pt}
  \hat g = \tilde g - \tilde g\, \tilde\beta^{-1}\, \tilde g\,
  \tilde\beta^{-1}\, \tilde g \;.
}

\end{itemize}


\subsection{Effective action for the superstring}

After having identified the Seiberg-Witten type relations \eqref{redefine} between
the frames  $(G,B)$ and $(\fa g,\fa\beta)$ for the fields in the gravity sector,
we now turn to the remaining massless
fields of the type II superstring.
Our guiding principle to construct an action for the latter is
that after the redefinition
\eqref{redefine}, the resulting action should be of the same form as before
with the usual objects
replaced by the corresponding ones in symplectic 
gravity, that is $\partial_a\rightarrow D^a$,
$H_{abc}\rightarrow \fa\Theta^{abc}$ etc.


\subsubsection*{The R-R sector}

Let us start with the Ramond-Ramond (R-R) sector and consider two sets of completely anti-symmetric fields $C_{a_1\ldots a_n}$ and $\fa C^{b_1\ldots b_n}$
in the frames $(G,B)$ and $(\fa g, \fa \beta)$, respectively.
As suggested by the result above, we can assume them to be related via
\eq{
\label{nicetrafo}
     C_{a_1\ldots a_n}=\fa\beta_{a_1 b_1}\ldots \fa\beta_{a_n b_n}\,
     \fa C^{b_1\ldots b_n}  \, .
}
Since $C_{a_1\ldots a_n}$ 
is invariant under $B$-field gauge transformations in the standard 
 frame, we know from the last subsection that 
 $\fa C^{b_1\ldots b_n}$ behaves as a $\beta$-tensor 
in the symplectic  frame.
Furthermore, using \eqref{covderivtrafo} we notice that also
\eq{
  \label{gen_potentials}
    \hat F^{a_1\ldots a_{n+1}}= \fa\nabla^{[a_1} \fa C^{a_2\ldots a_{n+1}]}
}
behaves as a $\beta$-tensor.
Finally, employing the first Bianchi identity of the Riemann tensor given in \eqref{bianchis}, we observe that \eqref{gen_potentials} is  invariant under gauge transformations
\eq{
  \delta_{\Lambda} \fa C^{a_1\ldots a_n} = \fa\nabla^{[a_1} \Lambda^{a_2 \ldots a_n]} \;,
}
and can thus be interpreted as a field strength.
Therefore, identifying $C_{a}$ and $C_{a_1 a_2 a_3}$ with the one- and three-form gauge potentials of type IIA supergravity, we have found corresponding expressions in the symplectic frame.

In analogy to the standard formulation, we then introduce generalized field strengths of the form
\eq{\label{IIBRR}
  \fa {\cal F}_2 =\fa F_2 \ , \hspace{40pt} \fa{\cal F}_4=\fa F_4-\fa\Theta\wedge \hat C_1\, ,
}
and for the corresponding action we consider
\eq{
\label{iiaactions}
\fa S^{\mbox{\scriptsize R-R}}_{\rm IIA}=\frac1{2\kappa_{10}^2}\int d^{10}x
\hspace{1pt}
\sqrt{- |\fa g|}\hspace{1pt}\bigl|\hat \beta^{-1}\bigr| \,
\Bigl(-{\textstyle {1\over 2}} \hspace{0.5pt} |\fa {\cal F}_2|^2-{\textstyle{1\over 2}}\hspace{0.5pt}|\fa
{\cal F}_4|^2\Bigr)\, ,
}
where we employ
\eq{
\label{formsquareabs}
|\fa{\mathcal F}_p|^2={1\over p!} \hspace{1pt} \fa {\mathcal F}_{a_1\dots a_p} \hspace{0.5pt}\fa {\mathcal F}^{a_1\dots a_p}\; .
}
As explained above, the quantities appearing in the parentheses in \eqref{iiaactions} are tensors with respect to $\beta$- (as well as usual) diffeomorphisms and so the full action \eqref{iiaactions} is invariant under $\beta$-diffeomorphisms.
The remaining part of the Ramond-Ramond sector is given by the Chern-Simons action, which in standard type IIA supergravity  takes the form
\eq{
S^{\rm CS}_{\rm IIA}&=\frac1{4\hspace{1pt}\kappa_{10}^2}\int H\wedge F_4\wedge
C_3\\
&=  \frac1{4\hspace{1pt}\kappa_{10}^2} \, \frac{1}{3! \hspace{1pt}4! \hspace{1pt}3!}
\int d^{10}x \; \epsilon^{a_1\ldots a_{10}}\,
H_{a_1 a_2 a_3}\hspace{1pt} F_{(4)a_4 a_5 a_6 a_7}\hspace{1pt} C_{(3)a_8 a_9 a_{10}} \;,
}
with $\epsilon^{a_1\ldots a_{10}}=\pm 1$ denoting the epsilon symbol. 
Note that $\epsilon^{a_1\ldots a_{10}}/\sqrt{-|G|}$ is
 a tensor under usual diffeomorphisms and is  invariant under $B$-field gauge transformations.
Applying then the change of fields \eqref{redefine} and keeping in mind our above discussion, we arrive at the following expression in the symplectic frame
\eq{
  \label{csaction}
\fa S^{\rm CS}_{\rm IIA}
=  \frac1{4\hspace{1pt}\kappa_{10}^2} \, \frac{1}{3! \hspace{1pt}4! \hspace{1pt}3!}
\int d^{10}x \hspace{1pt}\bigl|\hat \beta^{-1}\bigr| \,
   \epsilon_{b_1\ldots b_{10}}\,
    \fa\Theta^{b_1 b_2 b_3}\, \fa F_{(4)}^{b_4 b_5 b_6 b_7}\,
    \fa C_{(3)}^{b_8 b_9 b_{10}} \;.
} 
Note that from our remark around \eqref{nicetrafon} it follows that
$\epsilon_{b_1\ldots b_{10}}/\sqrt{-|\fa g|} $ transforms as a $\beta$-tensor, 
so that the Chern-Simons action is invariant under $\beta$-diffeomorphisms.

Finally, the gauge symmetries of the R-R fields  carry over from the usual case. In particular, as can be checked along similar lines compared to the standard situation, the  actions \eqref{iiaactions} and \eqref{csaction} are invariant under the following set of gauge transformations
\eq{
    &\delta_{\Lambda_{(0)}} \fa C^{ a} = \fa\nabla^{a} \Lambda_{(0)} \;,
    \hspace{90pt}
    \delta_{\Lambda_{(2)}} \fa C^{ a_1a_2 a_3} = \fa\nabla^{[a_1} \Lambda_{(2)}^{a_2  a_3]} \;, \\
    &\delta_{\Lambda_{(0)}} \fa C^{a_1 a_2 a_3} = - \Lambda_{(0)} \hspace{1pt}\fa \Theta^{a_1 a_2 a_3} \;.
}
In order to verify the invariance under $\delta_{\Lambda_{(0)}}$, 
the second Bianchi identity in \eqref{bianchi_plain} has to be employed, which can be brought into the form
\eq{
	\fa\nabla^{[a}\fa\Theta^{bcd]} = 0 \;.
}


\subsubsection*{The NS-R and R-NS sectors}

After having studied the bosonic part of the type IIA supergravity action,
we now turn to the part involving the 
gravitino $\Psi_a$ and the dilatino $\lambda$.
Let us  first establish our notation and  state that
\begin{equation}
  \nonumber
  \begin{array}{l@{\hspace{30pt}}l}
   \alpha, \beta, \gamma,\ldots      &\mbox{denote Lorentz-frame indices,} \\
   a,b,c,\ldots  &\mbox{denote space-time indices.}
   \end{array}
\end{equation}
The vielbein matrices $e_{\alpha}{}^a$ relating these two frames via $e_a = e_a{}^\alpha\,e_\alpha$ and $e^a = e_\alpha{}^a\,e^\alpha$ are  defined in the usual way by requiring that
\eq{
  \label{vielbein_04}
  e_\alpha{}^a \, e_\beta{}^b \hspace{1pt} G_{ab} = \eta_{\alpha\beta} \;,
}
with $\eta_{\alpha\beta} = \mbox{diag}\, (-1,+1,\ldots, +1)$. Therefore, our conventions are such that $e_\alpha{}^a\,e_a{}^\beta = \delta_\alpha^\beta$ and $e_\alpha{}^a\,e_b{}^\alpha = \delta_b^a$. 
The components of the spin connection $\omega^{\alpha}{}_{\beta}$ can be expressed in terms of the Christoffel symbols $\Gamma^c{}_{ab}$ in the following way
\eq{
  \label{spinchris}
  \omega_c{}^{\alpha}{}_{\beta}=e_a{}^{\alpha}{}\, e_{\beta}{}^b\, \Gamma^a{}_{cb}
  + e_a{}^{\alpha}{}\,  \partial_c   e_{\beta}{}^a\, ,
}
and for the gamma matrices we use the vielbein matrices to write $\gamma^a=\gamma^{\alpha} e_{\alpha}{}^a$. We furthermore define
$ \gamma^{a_1\ldots a_n}=\gamma^{[a_1}\,\gamma^{a_1}\, \ldots\, \gamma^{a_n]}$, so that
the kinetic term for the dilatino $\lambda$ can  be expressed as
\eq{
  \label{action_dilatini}
    {\cal L}^\lambda_{\rm IIA} =   \overline{\lambda}\hspace{1pt} \gamma^{a}\left(
            \partial_a - {\textstyle {i\over 4}} \hspace{1pt}
    \omega_{a\,\alpha\beta}\hspace{1pt} \gamma^{\alpha\beta}\right) \lambda \;,
}
where we have lowered the Lorentz-frame index of the spin connection with the metric $\eta_{\alpha\beta}$.

We now would like to obtain the action corresponding to \eqref{action_dilatini} in the non-geometric frame.
We therefore define the symplectic vielbein matrices $\fa e_a{}^{\alpha}$ for the metric $\fa g^{ab}$ by
\eq{
  \fa e^{\alpha}{}_a \, \fa e^{\beta}{}_b \hspace{1pt} \fa g^{ab} = \eta^{\alpha\beta} \;.
}
Comparing then with \eqref{vielbein_04}, we can infer that
\eq{
     \fa e^{\alpha}{}_a=\eta^{\alpha\beta}\, e_{\beta}{}^b\, \fa\beta_{ba}  \, .
}
For the transformation of the action \eqref{action_dilatini} under \eqref{redefine} we  employ
the relation \eqref{transgamma}, and we note again that  Lorentz-frame indices will be raised and lowered by $\eta_{\alpha\beta}$.
Using furthermore $\fa \gamma_a= \gamma_{\alpha} \fa e^{\alpha}{}_a$ and defining $\fa \lambda = \lambda$, we obtain
\eq{
    \fa{\cal L}^\lambda_{\rm IIA} =   \overline{\fa\lambda}\hspace{1pt} \fa\gamma_{a}\left(
            D^a - {\textstyle {i\over 4}} \hspace{1pt}
    \fa\omega^a{}_{\beta \delta} \gamma^{\beta\delta}\right) \fa\lambda \;,
}
with the symplectic spin-connection given by
\eq{
    \fa\omega^a{}_{\alpha}{}^{\beta}=\fa e_{\alpha}{}^b\, \fa e^{\beta}{}_c\, \fa\Gamma_b{}^{ac}
   + \hat e_{\alpha}{}^b  D^a \fa e^{\beta}{}_b\, .
}
The form of the kinetic term of the dilatino is thus preserved under the field redefinitions \eqref{redefine}.

A similar analysis can be performed for the kinetic term of the gravitino, 
which is given by the Rarita-Schwinger Lagrangian
\eq{ \label{RaritaSchwinger}
      {\cal L}^{\Psi}_{\rm IIA} =   \overline{\Psi}\vphantom{\fa\Psi}_a\hspace{1pt}
      \gamma^{abc} \left(
            \nabla_b -{\textstyle {i\over 4}} \hspace{1pt}
       \omega_{b\, \alpha\beta}\, \gamma^{\alpha\beta}\right) \Psi_c\, .
}
Here, we have written the covariant derivative $\nabla_a$ 
instead of the partial derivative, but
due to the anti-symmetrization in $\gamma^{abc}$ the  connection coefficients
drop out.
Now, to do the transformation into the symplectic frame, we first define
\eq{ 
\fa \Psi^a = \fa \beta^{ab} \hspace{1pt} \fa \Psi_b \; ,
}
where the additional factor of $\fa\beta^{ab}$ is crucial in order to cancel 
the inhomogenous term in the transformation of the Christoffel symbols
\eqref{transgamma}.
We then arrive  at a result
which is of the same form as \eqref{RaritaSchwinger}, namely
\eq{ \label{symplecticRaritaSchwinger}
      \fa{\cal L}^{\Psi}_{\rm IIA} =   \overline{\fa\Psi}\vphantom{\fa\Psi}^a\hspace{1pt}
      \fa\gamma_{abc} \left(
            \fa \nabla^b -{\textstyle {i\over 4}} \hspace{1pt}
      \fa\omega^b{}_{\alpha\beta}\, \gamma^{\alpha\beta}\right) \fa\Psi^c \;.
}
Again, the symmetric part of the connection does not contribute, but
the anti-symmetric part $\fa\Gamma_c{}^{[ab]}={1\over 2}{\cal
Q}_c{}^{ab}$ does appear in the symplectic Rarita-Schwinger action.

Finally, it can be checked that also the remaining fermionic terms in the type
IIA
supergravity Lagrangian transform as expected, and so we  arrive at a symplectic supergravity 
action. However, to study in detail the realization of supersymmetry, that is
the question of how
the bi-diffeomorphism invariance is extended
to supersymmetry, is beyond the scope of this paper.


\section{Solutions for non-geometric  backgrounds}
\label{sec_eom}

In section~\ref{sec_the_real_eom} we have derived the equations of motion \eqref{eom_final}
for the symplectic gravity theory \eqref{finalaction}, and in this section we are going to construct
solutions thereof.
In order to compare  these solutions to the ones obtained in generalized geometry and
double field theory, let us briefly recall our notation. In particular, there are three frames of 
fields we are going to employ in the following:
\eq{
\nonumber
\renewcommand{\arraystretch}{1.2}
\begin{array}{c@{\hspace{5pt}:\hspace{10pt}}l}
(G,B) & \mbox{standard geometric frame,} \\
(\fa g, \fa \beta) & \mbox{non-geometric frame obtained via the field redefinition \eqref{redef},} \\
(\fc g, \fc \beta) & \mbox{non-geometric frame obtained via the field redefinition \eqref{relation_dft}.} 
\end{array}
}
The standard frame can be formulated as a proper Lie algebroid defined on the tangent bundle together with the Lie bracket. Similarly, as we have discussed in the previous sections, also the non-geometric $(\fa g,\fa\beta)$-frame can  be expressed in terms of a proper Lie algebroid, which is defined on the co-tangent bundle endowed with the $H$-twisted Koszul bracket \eqref{koszulH}. In this sense, these two frames are distinguished.

Furthermore, by studying solutions to the equations of motion we can 
identify which frame provides
a natural description for what type of non-geometric background. 
To address this question, we proceed along two routes:
\begin{itemize}

\item First, we consider configurations T-dual to known solutions
of the field equations in the geometric $(G,B)$-frame. In fact, as mentioned in the introduction,
the approximate solution  of a flat torus  with constant $H$-flux
was the starting point from which the picture of non-geometric backgrounds
with $Q$- and $R$-flux has emerged \cite{Shelton:2005cf}.

\item Second, we can apply the field redefinition \eqref{redef} to  geometric solutions 
in the $(G,B)$-frame. Our expectation is that for generic $H$-fluxes the resulting
field configuration will have singularities and monodromies.
However, for instance  for Calabi-Yau backgrounds with vanishing $H$-flux, 
the latter problems can be absent.

\end{itemize}


\subsection{The constant $Q$-flux background}

We start by recalling  an approximate solution to the usual string equations
of motion in the $(G,B)$-frame (see also \cite{Halmagyi:2009te,Andriot:2011uh} and references therein).
The three-dimensional metric $G_{ab}$, the
dilaton $\phi$ and $B$-field are given by
\eq{ \label{standardbackground}
  G_{ab} = \delta_{ab} \;, \hspace{40pt}
  \phi = {\rm const.}\,, \hspace{40pt}
  B_{12} = 1 + h \hspace{1pt} x_3 \;,
}
which results in a constant $H$-flux.
As one can check, this ansatz only solves the field equations
in the standard setting up to terms linear  in $H$.
Applying successive T-dualities along the two directions
$x_1$ and $x_2$, one arrives at a background with metric
\eq{
\label{qfluxsol}
  G_{11}= G_{22} = \frac{1}{1+(1+h \hspace{1pt} x_3)^2} \;, \hspace{40pt}
  G_{33} = 1\;,
}
and non-vanishing $B$-field components
\eq{
  \label{qfluxsolB}
  B_{12} = - \frac{1+ h \hspace{1pt} x_3}{1+(1+h \hspace{1pt} x_3)^2} \;.
}
Note that this configuration is a so-called T-fold where the transition function between charts of the manifold have to include also T-duality transformations. Hence, as a geometric manifold it is not well-defined globally.

However, using the relation \eqref{relation_dft} for the metric  and 
$B$-field \eqref{qfluxsol} and \eqref{qfluxsolB}, respectively, 
in  the non-geometric $(\tilde g,\tilde\beta)$-frame 
we obtain a background of the form 
\eq{  
  \tilde g^{ab} = \delta^{ab} \;, \hspace{40pt}
  \tilde \beta^{12} = 1 + h \hspace{1pt} x_3 \;,
}
which means that here $\fc g^{ab}$ and $\fc \beta^{ab}$ are non-singular and thus
well-defined. The non-trivial components of the corresponding $Q$- and $R$-flux can then be computed as 
\eq{
  \tilde Q_3{}^{12}= h  \;,\hspace{40pt}
  \tilde\Theta^{123}=0 \;.
}  
Therefore, for describing this non-geometric background
with constant $\tilde Q$-flux, the
$(\tilde g,\tilde\beta)$-frame is appropriate.
This may have been expected because the relevant quantities in this setting 
are the fluxes $\tilde Q$ and $\tilde \Theta$.
On the other hand, since  in three dimensions the matrix $B_{ab}$ in \eqref{qfluxsolB} is not
invertible, the associated fields in the  non-geometric 
$(\fa g,\fa \beta)$-frame are not well-defined and therefore not suited to describe this configuration.

After formally applying a third T-duality  to the 
solution \eqref{qfluxsol} in the $x_3$-direction, it was argued that the resulting background is non-geometric with constant $R$-flux. 
The Buscher rules cannot be applied since  $x_3$  
is not a direction of isometry, and so
the explicit form of the background in the  $(G,B)$-frame is not known. 
But, in the  $(\fa g,\fa \beta)$-frame this configuration would correspond to having non-vanishing $\fa\Theta$ for vanishing $\fa{\cal Q}$-flux. Let us therefore directly analyze whether such a solution does exist.


\subsection{A solution with constant $R$-flux}

We note that because  the 
\mbox{(quasi-)}symplectic structure $\fa\beta^{ab}$ is invertible only in an even 
number of dimensions, we choose a flat four-dimensional metric $\fa g^{ab}=\delta^{ab}$
together with a constant dilaton. For the anti-symmetric  and invertible  bi-vector
we make the ansatz
\eq{ \label{solution+}
    \fa\beta=\left(\begin{matrix}
       0 & +\epsilon^{-1}\hspace{1pt} (1+x_4) & 0 & 0\\
     - \epsilon^{-1}\hspace{1pt} (1+x_4) & 0 & 0 & 0\\
     0 & 0 & 0 &+ \epsilon\hspace{1pt}\theta\\
     0 & 0 &  -\epsilon\hspace{1pt} \theta & 0\end{matrix}\right)
     \hspace{30pt}
      {\rm for}\ x_4>0 \;,
}
and
\eq{ \label{solution-}
   \fa\beta=\left(\begin{matrix}
       0 &  +\epsilon^{-1}\hspace{1pt} (1-x_4) & 0 & 0\\
     - \epsilon^{-1}\hspace{1pt} (1-x_4) & 0 & 0 & 0\\
    0 & 0 & 0 & -\epsilon\hspace{1pt}\theta\\
    0 & 0 &  +\epsilon\hspace{1pt} \theta & 0\end{matrix}\right)
    \hspace{30pt}
    {\rm for}\ x_4<0 \;,
}
with constant parameters $\epsilon$ and $\theta$. 
Since we are considering the non-compact space $\mathbb{R}^4$, we have chosen 
two patches to avoid zeros of $\fa\beta$ and
singularities of $\fa{\cal Q}$ and $\fa R^{ab}$.
The only non-trivial component of the $R$-flux
following from this ansatz is found as
\eq{
  \fa\Theta^{123} = \theta \;.
}
To compute the Ricci tensor, we first determine the non-vanishing components of the $\fa{\cal Q}$-flux as
\eq{
    \fa{\cal Q}_1{}^{31}=-\fa{\cal Q}_1{}^{13}=\fa{\cal Q}_2{}^{32}=-\fa{\cal
       Q}_2{}^{23}=  {\theta\hspace{1pt}\epsilon\over  1+|x_4|}\, .
}
The non-vanishing components
of the Levi-Civita connection \eqref{klc} are then given by
\eq{
     \fa \Gamma_3{}^{11}
   =\fa \Gamma_3{}^{22}
   =-\fa \Gamma_1{}^{13}
   =-\fa \Gamma_2{}^{23}
   = {\theta\hspace{1pt}\epsilon\over  1+|x_4|}\, ,
}
leading to the following  components
of the Ricci-tensor $\fa R^{ab}$
\eq{
      \fa R^{11}= \fa R^{22}= \frac{3}{4}\, \fa R^{33}=-
     3 \, {(\theta\hspace{1pt}\epsilon)^2\over  (1+|x_4|)^2} \;.
}
Hence,  the Ricci tensor does not vanish identically and so
the field equations \eqref{eom_final} are not satisfied (up to linear order in the flux).
However, in the limit $\epsilon\to 0$ the components of $\fa R^{ab}$
as well as the $\fa{\cal Q}$-flux
approach zero for each value of $x_4$,
while the $R$-flux $\fa\Theta$ remains constant
\eq{
  \fa R^{ab} \xrightarrow{\;\;\epsilon\to 0 \;\;} 0 \;, \hspace{40pt}
  \fa {\mathcal Q}_c{}^{ab} \xrightarrow{\;\;\epsilon\to 0 \;\;} 0 \;, \hspace{40pt}  
  \fa \Theta^{123}=\theta \;.
}  
Note that in contrast to $\fa{\mathcal Q}$, the flux $\fa Q_4{}^{12}$ is not well-defined in this limit,
which agrees with our observation that the appropriate object in the $(\fa g,\fa\beta)$-frame is the $\fa{\cal Q}$-flux.

Before closing this section, let us study how the solution presented here maps
to the geometric $(G,B)$-frame. In particular, applying the transformation
\eqref{redef} to the bi-vector $\fa\beta$ specified by \eqref{solution+} and
\eqref{solution-}, we arrive at the following form of the $B$-field components
in the patch $x_4>0$
\eq{
B_{12} =\, -\frac{\epsilon}{1+x_4}\;, \hspace{40pt}
B_{34}=\,-\frac{1}{\epsilon \hspace{1pt}\theta}  \;.
}
For the metric we obtain
\eq{ 
G_{11} = G_{22} = \frac{\epsilon^2}{(1+x_4)^2} \;, \hspace{40pt}
G_{33}=G_{44}= \frac{1}{(\epsilon \hspace{1pt}\theta)^2} \;,
}
and therefore, although the corresponding $H$-flux behaves properly, the metric is ill-defined in the limit $\epsilon\to0$, even locally. 
We  conclude that for describing solutions of this type
(having constant $R$-flux), the $(\fa g,\fa \beta)$-frame is suitable.


\subsection{Calabi-Yau manifolds in the non-geometric frame}

Our guiding principle for finding the solutions in the last two sections
was T-duality. But, another way of generating configurations which solve
the field equations  \eqref{eom_final}
is  to directly transform  from the geometric  
to the non-geometric frame via the field redefinition \eqref{redefine}.
For a general non-trivial
$B$-field one would expect that the transformed solution 
has monodromies, which means that 
 the non-geometric frame is not suited to describe such configurations.
However, for vanishing $H$-flux we expect
that solutions can directly be transformed
using \eqref{redefine} without encountering such problems.

A large set of  solutions to string theory with vanishing $H$-flux are Calabi-Yau manifolds, 
which are complex manifolds satisfying
\eq{
\arraycolsep2pt
  \begin{array}{lclcrcl}
      R_{ab}&=&0\;, &\hspace{40pt} &   d\omega&=&0 \;, \\[5pt]
      H_{abc}&=&0\;, &&\phi&=&{\rm const.} \;,
    \end{array}
}
where $\omega$ denotes the K\"ahler form $\omega={i\over 2} \hspace{1pt}G_{a\overline b}\, dz^a\wedge
d\overline{z}^{\overline b}$ in complex coordinates.
Choosing then the following bi-vector  for a complex three-manifold
\eq{
    \fa\beta=\left(\begin{matrix}
       0 & +1 & 0 & 0 & 0 & 0\\
     - 1 & 0 & 0 & 0& 0 & 0 \\
     0 & 0 & 0 & +1 & 0 & 0\\
     0 & 0 &  -1& 0 & 0 & 0\\
   0 & 0 & 0 & 0 & 0 & +1\\
     0 & 0 &  0& 0 & -1 & 0
\end{matrix}\right) \;,
}
and using the field redefinition \eqref{redefine}, we obtain a smooth solution to the 
string equations of motion in the non-geometric frame
 characterized by
\eq{
      \fa R_{ab}=0  \;,\hspace{40pt}
      \fa\Theta_{abc}=0\;,\hspace{40pt}
       \phi={\rm const.} 
}
The K\"ahler form $\omega$ is mapped from the geometric to the non-geometric frame as
\eq{
W = \frac{i}{2}\,\fa g^{a\bar{b}}\,\partial_{z_a}\wedge\partial_{\bar{z}_b}\;,
}      
and in the following we want to ask which conditions this two-vector has to satisfy in order for the 
resulting manifold to again be a Calabi-Yau manifold.

Let us therefore first note that the exterior derivative in the framework of the $H$-twisted
Koszul bracket is characterized by \eqref{algebroiddiff} and  reads
\eq{
	d_\beta^H \alpha(\xi_0, \dots, \xi_n) = 
	\hspace{10pt}& \sum_{i=0}^n \;(-1)^i 
		(\beta^\sharp\xi_i)\,\alpha(\xi_0,\dots,\hat{\xi}_i,\dots,\xi_n)  \\
	+&\sum_{i<j} \; (-1)^{i+j} 
	\alpha \bigl([\xi_i, \xi_j]_K^H,\xi_0,\dots,
		\hat{\xi}_i,\dots,\hat{\xi}_j,\dots,\xi_n \bigr) \;,
}
where $\alpha\in\Gamma(\wedge^nTM)$. 
Employing the Jacobi-identity \eqref{jacHK} of $[\,\cdot\,,\,\cdot\,]_K^H$ for the $R$-flux \eqref{RaH}, we see that $d_\beta^H$ is nilpotent, that is
\eq{
  \bigl(d_\beta^H\bigr)^2=0 \;.
}
This allows us to define a \emph{quasi-Poisson cohomology} by considering the quotient of $d_\beta^H$-closed forms by $d_\beta^H$-exact forms.
Next, recalling \eqref{algebroiddiff} also for the Lie bracket
and similarly as in \eqref{anchG}, for a general $n$-form $\rho$
we compute
\eq{\label{dRtoP}
	\bigl(\wedge^{n+1}\beta^\sharp\,d\rho \bigr)(\xi_0, \ldots,\xi_n)
	&= (-1)^{n+1}\,d\rho\bigl(\beta^\sharp\xi_0, \ldots,\beta^\sharp\xi_n\bigr) \\
	&= -\bigl(d_\beta^H(\wedge^n\beta^\sharp\,\rho)\bigr)(\xi_0,...,\xi_n) \, ,
}
where we used that $\beta$ is an algebra homomorphism, i.e. $\beta^\sharp[\xi_i,\xi_j]_K^H = [\beta^\sharp\xi_i,\beta^\sharp\xi_j]$. 
Therefore,   a symplectic form $\omega$ can be translated to a $d_\beta^H$-closed two-vector field $W$ as follows
\eq{
	d\omega=0 \qquad\Longrightarrow\qquad
	d_\beta^H\bigl( \wedge^2\beta^\sharp\omega \bigr) \equiv d_\beta^H W = 0 \;,
} 
where we 
have identified  $W$ as the analogue of $\omega$.
Finally, since $\beta^\sharp$ is assumed to be bijective, $W$ is  non-degenerate if $\omega$ has that property. Thus, coming back to the beginning of this section, we have shown that under the field redefinition \eqref{redefine} a Calabi-Yau manifold is mapped to a space with very similar properties.
We define the latter as follows
\begin{itemize}

\item[]{\bf Definition:} A \emph{co-Calabi-Yau manifold} is a complex manifold admitting a non-degenerate $d_\beta^H$-closed two-vector field $W$ and an associated hermitean metric $\fa g$ characterized by
\eq{
	W = \frac{i}{2}\,\fa g^{a\bar{b}}\,\partial_{z_a}\wedge\partial_{\bar{z}_b}\,,
}
for which the Ricci tensor $\fa R^{ab}$ vanishes. 

\end{itemize}

Our findings in this section then imply that if a Calabi-Yau manifold is a 
solution to the equations of motion in the 
geometric frame (with vanishing $H$-flux and constant dilaton), then 
there exists  a corresponding co-Calabi-Yau manifold
(with   vanishing $R$-flux and constant dilaton)
which is a  solution to the field 
equations \eqref{eom_final} in the symplectic frame.


\subsection*{Remarks}

After having discussed two different types of solutions to the equations of motion of our symplectic gravity theory, we can give  the following conceptual interpretation.
\begin{itemize}

\item The theory characterized by the symplectic gravity action
\eqref{finalaction_pre}  provides
an  effective  {\it field theory} description of the
deep non-geometric (world-sheet asymmetric) regime of string
theory, which is well-suited for backgrounds with
non-vanishing $R$- and ${\cal Q}$-flux.

\item As it was mentioned already in the beginning of this section,
from the point of view of double field theory
there seem to exist (at least) two distinguished frames in which
the action and the equations of motion take the familiar form.
These are the geometric frame characterized by the fields $(G_{ab},B_{ab},\phi)$ and
the non-geometric frame with fields
$(\hat g^{ab},\fa\beta^{ab},\phi)$. 
From a mathematical point of view, these 
are frames in which the Courant algebroid 
reduces to  proper Lie algebroids 
on $TM$ and $T^*M$, respectively. As a 
consequence, one  can perform constructions similar as in standard differential geometry. 
An immediate question which arises then is, if there are other frames 
allowing for such constructions as well, possible with different fluxes turned on, and whether
there exists a classification thereof.

\item We expect that the symplectic-gravity equations of motions 
    should also admit solutions describing the complete T-dual 
   of the fundamental string and the NS five-brane. From a world-sheet 
  perspective these might also be called solutions for the 
  asymmetric string and asymmetric NS five-brane. 
 Contrarily, as the supergravity solutions for D-branes have vanishing
  $H$-flux, these can simply be transformed from the $(G,B)$ to the 
   $(\fa g,\fa \beta)$-frame without
  any obstacles. Thus, in this sense they are similar to
  the Calabi-Yau solutions.  

\end{itemize}


\section{Conclusions}
\label{sec_concl}

The objective of this paper was to formulate an effective field
theory description of the deep non-geometric regime of string
theory, where the $R$-flux is non-vanishing. 
Our initial intuition was that there should exist a framework
which closely resembles the one of  standard Riemannian geometry,
in which the string effective actions are usually described.
It turned out that the differential geometry based
on the  theory of Lie algebroids indeed serves this purpose.

Here, in contrast to the usual setting of a manifold
equipped with a metric on the tangent bundle, we considered the co-tangent bundle 
endowed with a metric $\fa g^{ab}$ and a \mbox{(quasi-)}symplectic structure
$\beta^{ab}$.
Employing results from the mathematics literature, it was possible 
to construct a differential geometry for our Lie algebroid,
which guarantees that all geometric objects are covariant with respect
to standard diffeomorphisms.
However, it is quite intriguing that an additional symmetry could
be identified so that the geometry
is  not only covariant with respect to standard diffeomorphism, but
also with respect to these  so-called $\beta$-diffeomorphisms.
This second local symmetry is  emanating from the gauge symmetry
of the Kalb-Ramond field in the geometric frame.

Based on this symplectic generalization of differential geometry,
it was straightforward to construct
a bi-invariant Einstein-Hilbert type action for the
dynamical fields, i.e. the metric, the (quasi-)symplectic structure
and the dilaton. Remarkably, this action
is of the same form as the usual effective action for the bosonic string
in the geometric frame.
Since the (quasi-)symplectic form implicitly appeared
even in the derivative,
it was a non-trivial exercise to show
that even the equations of motion are of the same form.
Here, the appearance of the determinant of the (quasi-)symplectic
form in the integration measure was crucial.

We showed
that the fields in the geometric frame and in the non-geometric
one are related via a field redefinition, whose form
is reminiscent of the Seiberg-Witten map, which appeared
in the context of open strings in two-form backgrounds.
Employing  this map, we first showed explicitly that the 
symplectic gravity action
is directly related to the action for the gravitational sector of 
string theory. 
Furthermore, we applied and extended this field redefinition also to 
the Ramond-Ramond sector
and to the fermionic terms, which allowed us to propose a 
symplectic supergravity action. 
To really extend
our bi-symmetry principle to the supersymmetric case was
beyond the scope of this paper, but it is certainly
an important step to be carried out in the future.
Additionally, we pointed out that the field redefinition straightforwardly
allows to transform higher-order $\alpha'$-corrections, hence  leading
to their form in the non-geometric frame of string theory.

Finally, we studied solutions to the equations of motion in the non-geometric
frame.
In particular, we considered two flat  backgrounds with constant 
$Q$- and $R$-flux as well as the symplectic analogue of Calabi-Yau manifolds. 

Clearly, as mentioned in the course of this paper, 
there are many open questions and directions 
worth to be   studied  in more detail in the future. 
Specifically, we would like to mention
that the \mbox{(quasi-)}symplectic structure $\fa\beta^{ab}$ 
naturally defines  a \mbox{(quasi-)}Poisson
structure. One may therefore speculate that the theory developed
here can be considered as the classical limit of a quantum-deformed 
symplectic gravity theory. This would  serve as
a concrete mathematical realization of the idea of non-associative gravity, as
it was verbally proposed in \cite{Blumenhagen:2010hj}.


\vspace*{1.25cm}

\subsubsection*{Acknowledgments}

We thank David Andriot, Dieter L\"ust, Peter Patalong and  Dan Waldram for discussion and  Christian Schmid for comments on the manuscript. R.B. and F.R. thank the Simons Center for Geometry and Physics for hospitality. 
E.P. is supported by the Padova University Project CPDA105015/10
and by the MIUR-FIRB grant RBFR10QS5J.


\clearpage
\bibliography{references}
\bibliographystyle{utphys}


\end{document}